\documentclass[12pt]{article}

\usepackage{axodraw}

\makeatletter
\if@twoside \oddsidemargin -17pt \evensidemargin 00pt
\else \oddsidemargin 00pt \evensidemargin 00pt
\fi
\expandafter\ifx\csname mathrm\endcsname\relax\def\mathrm#1{{\rm #1}}\fi

\renewcommand{\theequation}{\thesection.\arabic{equation}}
\newcounter{saveeqn}

\@addtoreset{equation}{section}

\makeatother

\def\co{\relax}
\def\co{,}
\def\nl{\nonumber\\}
\def\nlc{\co\nonumber\\}
\def\nln{\nonumber\\*[-1ex]\phantom{\fbox{\rule{0em}{2ex}}}}

\def\beq{\begin{equation}}
\def\eeq{\end{equation}}
\def\beqar{\begin{eqnarray}}
\def\eeqar{\end{eqnarray}}
\def\barr#1{\begin{array}{#1}}
\def\earr{\end{array}}
\def\bfi{\begin{figure}}
\def\efi{\end{figure}}
\def\btab{\begin{table}}
\def\etab{\end{table}}
\def\bce{\begin{center}}
\def\ece{\end{center}}
\def\nn{\nonumber}
\def\disp{\displaystyle}
\def\text{\textstyle}

\def\al{\alpha}
\def\be{\beta}
\def\ga{\gamma}
\def\de{\delta}
\def\veps{\varepsilon}
\def\la{\lambda}
\def\si{\sigma}
\def\Ga{\Gamma}
\def\De{\Delta}
\def\La{\Lambda}

\def\refeq#1{\mbox{(\ref{#1})}}
\def\reffi#1{\mbox{Fig.~\ref{#1}}}
\def\reffis#1{\mbox{Figs.~\ref{#1}}}
\def\refta#1{\mbox{Table~\ref{#1}}}
\def\reftas#1{\mbox{Tables~\ref{#1}}}
\def\refse#1{\mbox{Sect.~\ref{#1}}}

\def\refapp#1{\mbox{Appendix~\ref{#1}}}
\def\citere#1{\mbox{Ref.~\cite{#1}}}
\def\citeres#1{\mbox{Refs.~\cite{#1}}}

\def\solid{\raise.9mm\hbox{\protect\rule{1.1cm}{.2mm}}}
\def\dash{\raise.9mm\hbox{\protect\rule{2mm}{.2mm}}\hspace*{1mm}}

\newcommand{\GeV}{\unskip\,\mathrm{GeV}}

\newcommand{\pba}{\unskip\,\mathrm{pb}}

\def\mathswitchr#1{\relax\ifmmode{\mathrm{#1}}\else$\mathrm{#1}$\fi}

\newcommand{\PW}{\mathswitchr W}
\newcommand{\PZ}{\mathswitchr Z}

\newcommand{\Pe}{\mathswitchr e}
\newcommand{\Pne}{\mathswitch \nu_{\mathrm{e}}}
\newcommand{\Pane}{\mathswitch \bar\nu_{\mathrm{e}}}

\newcommand{\Pep}{\mathswitchr {e^+}}
\newcommand{\Pem}{\mathswitchr {e^-}}

\newcommand{\PWp}{\mathswitchr {W^+}}
\newcommand{\PWm}{\mathswitchr {W^-}}
\newcommand{\PWpm}{\mathswitchr {W^\pm}}

\def\mathswitch#1{\relax\ifmmode#1\else$#1$\fi}

\newcommand{\MW}{\mathswitch {M_\PW}}

\newcommand{\MZ}{\mathswitch {M_\PZ}}

\newcommand{\Me}{\mathswitch {m_\Pe}}

\newcommand{\GW}{\mathswitch {\Gamma_\PW}}

\def\ie{i.e.\ }

\renewcommand{\O}{{\cal O}}
\newcommand{\C}{{\cal C}}
\newcommand{\Oa}{\mathswitch{{\cal{O}}(\alpha)}}

\newcommand{\ri}{{\mathrm{i}}}
\newcommand{\ieps}{\ri\epsilon}
\newcommand{\rd}{{\mathrm{d}}}

\newcommand{\M}{{\cal {M}}}

\newcommand{\F}{{\cal {F}}}
\newcommand{\bq}{{\bf q}}

\newcommand{\born}{{\mathrm{Born}}}
\newcommand{\nf}{{\mathrm{nf}}}

\def\Li{\mathop{\mathrm{Li}_2}\nolimits}
\def\cLi{\mathop{{\cal L}i_2}\nolimits}
\def\Re{\mathop{\mathrm{Re}}\nolimits}
\def\Im{\mathop{\mathrm{Im}}\nolimits}

\def\arc{\mathop{\mathrm{arc}}\nolimits}

\def\draftdate{\relax}
\def\mda{\relax}
\def\mua{\relax}
\def\mla{\relax}
\def\draft{
\def\thtystars{******************************}
\def\sixtystars{\thtystars\thtystars}
\typeout{}
\typeout{\sixtystars**}
\typeout{* Draft mode!
         For final version remove \protect\draft\space in source file *}
\typeout{\sixtystars**}
\typeout{}
\def\draftdate{\today}
\def\mua{\marginpar[\boldmath\hfil$\uparrow$]%
                   {\boldmath$\uparrow$\hfil}%
                    \typeout{marginpar: $\uparrow$}\ignorespaces}
\def\mda{\marginpar[\boldmath\hfil$\downarrow$]%
                   {\boldmath$\downarrow$\hfil}%
                    \typeout{marginpar: $\downarrow$}\ignorespaces}
\def\mla{\marginpar[\boldmath\hfil$\rightarrow$]%
                   {\boldmath$\leftarrow $\hfil}%
                    \typeout{marginpar: $\leftrightarrow$}\ignorespaces}
\def\Mua{\marginpar[\boldmath\hfil$\Uparrow$]%
                   {\boldmath$\Uparrow$\hfil}%
                    \typeout{marginpar: $\Uparrow$}\ignorespaces}
\def\Mda{\marginpar[\boldmath\hfil$\Downarrow$]%
                   {\boldmath$\Downarrow$\hfil}%
                    \typeout{marginpar: $\Downarrow$}\ignorespaces}
\def\Mla{\marginpar[\boldmath\hfil$\Rightarrow$]%
                   {\boldmath$\Leftarrow $\hfil}%
                    \typeout{marginpar: $\Leftrightarrow$}\ignorespaces}
\overfullrule 5pt
\oddsidemargin -15mm
\marginparwidth 29mm
}

\makeatletter

\def\eqnarray{\stepcounter{equation}\let\@currentlabel=\theequation
\global\@eqnswtrue
\global\@eqcnt\z@\tabskip\@centering\let\\=\@eqncr
$$\halign to \displaywidth\bgroup\hskip\@centering
  $\displaystyle\tabskip\z@{##}$\@eqnsel&\global\@eqcnt\@ne
  \hskip 2\arraycolsep \hfil${##}$\hfil
  &\global\@eqcnt\tw@ \hskip 2\arraycolsep $\displaystyle\tabskip\z@{##}$\hfil
   \tabskip\@centering&\llap{##}\tabskip\z@\cr}
\def\appendix{\par
 \setcounter{section}{0} \setcounter{subsection}{0}
 \def\thesection{\Alph{section}}}

\makeatother

\newcommand{\lsim}
{\;\raisebox{-.3em}{$\stackrel{\displaystyle <}{\sim}$}\;}
\newcommand{\gsim}
{\;\raisebox{-.3em}{$\stackrel{\displaystyle >}{\sim}$}\;}

\newcommand{\eeffff}{\Pep\Pem\to4\,\mbox{fermions}}
\newcommand{\eeWW}{{\Pe^+ \Pe^-\to \PW^+ \PW^-}}
\newcommand{\Wpff}{{\PW^+ \to f_1\bar f_2}}
\newcommand{\Wmff}{{\PW^- \to f_3\bar f_4}}

\newcommand{\Kp}{K_+}
\newcommand{\Km}{K_-^*}
\newcommand{\betaM}{\beta}
\newcommand{\betap}{\bar{\beta}}
\newcommand{\betaW}{\beta_\PW}

\newcommand{\xW}{x_\PW}
\newcommand{\Cbr}{{\cal C}}
\newcommand{\Dbr}{{\cal D}}
\newcommand{\Ebr}{{\cal E}}
\newcommand{\Ybr}{Y'}
\newcommand{\Ibr}{{\cal I}}
\newcommand{\CM}{\mathrm{CM}}
\newcommand{\real}{{\mathrm{real}}}
\newcommand{\virt}{{\mathrm{virt}}}
\newcommand{\ffp}{\mathswitch{\mathrm{f\/f}'}}
\newcommand{\mfp}{\mathswitch{\mathrm{mf}'}}
\newcommand{\mmp}{\mathswitch{\mathrm{mm}'}}

\begin{document}

\thispagestyle{empty}
\def\thefootnote{\fnsymbol{footnote}}
\setcounter{footnote}{1}
\null
\draftdate\hfill CERN-TH/97-258 \\
\strut\hfill PSI-PR-97-30\\
\strut\hfill hep-ph/9710521
\vskip 0cm
\vfill
\begin{center}
{\Large \bf
Non-factorizable photonic corrections to
\boldmath{$\Pep\Pem\to\PW\PW\to 4\,$fermions}%
\par} \vskip 2.5em
{\large
{\sc A.~Denner%
}\\[1ex]
{\normalsize \it Paul-Scherrer-Institut, W\"urenlingen und Villigen\\
CH-5232 Villigen PSI, Switzerland}\\[2ex]
{\sc S.~Dittmaier%
}\\[1ex]
{\normalsize \it Theory Division, CERN\\
CH-1211 Geneva 23, Switzerland}\\[2ex]
{\sc M. Roth%
}\\[1ex]
{\normalsize \it Paul-Scherrer-Institut, W\"urenlingen und Villigen\\
CH-5232 Villigen PSI, Switzerland\\
and\\
Institut f\"ur Theoretische Physik, ETH-H\"onggerberg\\
CH-8093 Zurich, Switzerland
}\\[2ex]
}
\par \vskip 1em
\end{center}\par
\vskip .0cm 
\vfill 
{\bf Abstract:} \par 
We study the non-factorizable corrections to $\PW$-pair-mediated
four-fermion production in $\Pep\Pem$ annihilation in double-pole
approximation. We show how these corrections can be combined with the
known corrections to the production and the decay of on-shell W-boson pairs,
and how the full off-shell Coulomb singularity is included.
Moreover, we find that the actual form of the real non-factorizable
corrections depends on the parametrization of phase space, more
precisely, on the definition of the invariant masses of the resonant 
\PW~bosons.
For the usual parametrization the full analytical results for the 
non-factorizable corrections are presented.
Our analytical and numerical results for the non-factorizable corrections 
agree with a recent calculation, which was found to differ from a 
previous one. The detailed numerical discussion covers 
the invariant-mass distribution, various angular distributions, and the 
lepton-energy distribution for leptonic final states.
\par
\vskip 1cm
\noindent
October 1997 
\par
\null
\setcounter{page}{0}
\clearpage
\def\thefootnote{\arabic{footnote}}
\setcounter{footnote}{0}

\section{Introduction}

After the successful experimental investigation of the $\PZ$~boson at
LEP1, at present the $\PW$~boson is studied via its pair production at
the upgraded collider LEP2. Since $\PW$~bosons decay very rapidly, the
actual reaction observed is $\eeffff$. The virtual $\PW$~bosons can be
reconstructed by selecting those events in which the invariant masses
of the final-state fermion pairs are close to the \PW-boson mass. The
expected integrated luminosity of LEP2 is $500\pba^{-1}$
\cite{LEP2EL97}. Accordingly, one expects to produce about 8000
$\PW$~pairs, and the typical experimental accuracy is of the order of
one to a few per cent. For the measurement of the $\PW$-boson mass the
expected accuracy is even below the per-mille level \cite{LEP2MW97}.

This experimental accuracy should be matched or better exceeded by the
precision of the theoretical predictions.  The calculation of the
cross section for $\eeffff$ with an accuracy of 1\% or better is a
non-trivial task.  In the LEP2 energy region, the lowest-order cross
section is dominated by the diagrams that involve two resonant
$\PW$~bosons. All other lowest-order diagrams are typically suppressed
by a factor $\Ga_\PW/\MW$, but may be enhanced in certain regions of
phase space. All these contributions are required at the 1\% level
(see \citere{LEP2Wevgen} and references therein).  At the level of
$\Oa$ corrections, on the other hand, it is sufficient to take into
account those contributions that are enhanced by two resonant
$\PW$~propagators (doubly-resonant contributions). These can be split
in two types: factorizable and non-factorizable corrections
\cite{Ae94}.  The factorizable corrections are those that correspond
either to the production of the \PW-boson pair or to the decay of one
of the \PW~bosons.  The corrections to on-shell \PW-pair production
\cite{rcwprod} and decay \cite{rcwdecay} are available.  The
non-factorizable corrections are those in which the production
subprocess and the decay subprocesses are not independent; typically
an additional particle is exchanged between two of the subprocesses.
Simple power-counting reveals that among the non-factorizable
corrections only those Feynman diagrams contribute to the
doubly-resonant contributions in which a soft photon is exchanged
between the subprocesses.

In this paper we explicitly calculate these non-factorizable photonic
corrections. It has been shown that they vanish in inclusive
quantities, \ie if the invariant masses of both $\PW$~bosons are
integrated out \cite{Fa94}. However, for non-inclusive quantities
these corrections do not vanish in general.  The non-factorizable
photonic corrections have already been investigated by two groups.
Melnikov and Yakovlev \cite{Me96} have given the analytical results
only in an implicit form and restrict the numerical evaluation to a
special phase-space configuration. Beenakker, Berends and Chapovsky
have provided both the complete formulae and an adequate numerical
evaluation \cite{Be97a,Be97b}, but do not find agreement with all
results of \citere{Me96}. For this reason, it is worth-while to
present the results of a third independent calculation.

We start out by discussing the definition of the virtual and real photonic
non-factoriz\-able corrections in double-pole approximation in detail.
Since only soft photons are relevant in double-pole approximation, 
the virtual non-factoriz\-able correction is just a factor to the 
lowest-order cross section. For the corresponding real correction, the
situation is similar, but in addition an integration over the photon 
momentum has to be performed. This requires a specification of the
phase-space parametrization,
which includes, in particular, the invariant masses of the \PW~bosons.
Usually these are defined via 
the invariant masses of the respective final-state fermion pairs
and are chosen as independent variables
\cite{Me96,Be97a,Be97b}. Experimentally, however,
the invariant mass of a \PW~boson is identified with the
invariant mass of the associated jet pair that necessarily includes
soft and collinear photons. Therefore,
the influence of the choice for the invariant masses of the \PW~bosons
on the non-factorizable corrections should be investigated 
in order to provide sound predictions for physical situations.

Besides the non-factorizable doubly-resonant corrections, the most
important effect of the instability of the \PW~bosons is the
modification of the Coulomb singularity. Since the off-shell Coulomb
singularity results from a scalar integral that also contributes to
the doubly-resonant non-factorizable corrections, it seems to be
natural to approximate this integral in such a way that both effects
are simultaneously included. This requires going beyond the strict
double-pole approximation.

The paper is organized as follows: in \refse{sedefapp} we discuss the
basis of the double-pole approximation, with particular emphasis on
the real corrections, and formulate a general gauge-independent
definition of the non-factorizable doubly-resonant corrections. In
\refse{secalc} we sketch the calculation of the scalar integrals and
describe the evaluation of the 5-point functions in detail. Section
\ref{seanres} summarizes our analytic results for the correction
factor; in particular, the cancellations between virtual and real
photonic corrections are inspected, and the off-shell Coulomb
singularity is embedded in the approach. Section \ref{senumres} is
devoted to the discussion of numerical results. The appendices provide
more information on the evaluation of the real bremsstrahlung
integrals and the explicit results for the required scalar integrals.

\section{Definition of the approximation}
\label{sedefapp}

\subsection{Conventions and notations}

In this paper we discuss corrections to the process
\beq\label{process}
\Pep(p_+) + \Pem(p_-) \;\to\; \PWp(k_+) + \PWm(k_-) \;\to\; 
f_1(k_1) + \bar f_2(k_2) + f_3(k_3) + \bar f_4(k_4).
\eeq
The relative charges of the 
fermions $f_i$ are represented by $Q_i$ with $i=1,\ldots,4$.
The masses of the external fermions, $m_i^2=k_i^2$ and
$\Me^2=p_\pm^2$, are neglected,
except where this would lead to mass singularities. 
The momenta of the intermediate \PW~bosons are defined by
\beq
k_+ = k_1+k_2, \qquad k_- = k_3+k_4,
\eeq
their complex mass squared and their respective invariant masses are denoted
by
\beq
M^2 = \MW^2-\ri\MW\Gamma_\PW, \qquad
M_\pm = \sqrt{k_\pm^2},
\eeq
respectively, and we introduce the variables
\beq
K_+ = k_+^2-M^2, \qquad K_- = k_-^2-M^2.
\eeq
Furthermore, we define the following kinematical invariants
\beqar
s &=& (p_+ + p_-)^2 = (k_+ + k_-)^2, \nl
s_{ij}  &=& (k_i + k_j)^2, \nl
s_{ijk} &=& (k_i + k_j + k_k)^2,
\qquad i,j,k=1,2,3,4, 
\eeqar
which obey the relations
\beqar
s &=& k_+^2 + k_-^2 + s_{13} + s_{14} + s_{23} + s_{24},
\qquad s_{12}=k_+^2, \qquad s_{34}=k_-^2, \nl
s_{ijk} &=& s_{ij} + s_{ik} + s_{jk},
\qquad i,j,k=1,2,3,4.
\eeqar

\subsection{Doubly-resonant virtual corrections}

The aim of the present paper is to evaluate the non-factorizable
corrections to the process \refeq{process} in 
double-pole approximation (DPA). The DPA takes
into account only the leading terms in an expansion around the poles
originating from the two resonant \PW~propagators.

In DPA, the lowest-order matrix element for the process
\refeq{process} factorizes into the matrix element for the production
of the two on-shell \PW~bosons, $\M^\eeWW_\born(p_+,p_-,k_+,k_-)$, the
(transverse parts of the) propagators of these bosons, and the matrix
elements for the decays of these on-shell bosons,
$\M^\Wpff_\born(k_+,k_1,k_2)$ and $\M^\Wmff_\born(k_-,k_3,k_4)$:
\beq\label{mborn}
{\cal M}_{\born} = 
\sum_{\la_+,\la_-}\frac{\M^\eeWW_\born \M^\Wpff_\born \M^\Wmff_\born}{K_+ K_-}.
\eeq
The sum runs over the physical polarizations $\la_\pm$ of the 
\PWpm~bosons. 

The higher-order corrections to \refeq{process} can be separated into
factorizable and non-factorizable contributions \cite{Ae94}.  In the
factorizable contributions the production of two $\PW$~bosons and
their subsequent decays are independent. The corresponding Feynman
diagrams can be split into three parts by cutting only the two
\PW-boson lines. The corresponding matrix element factorizes in the
same way as the lowest-order matrix element \refeq{mborn}.

The non-factorizable corrections comprise all those 
contributions in which W-pair
production and/or the subsequent W decays are not independent.
Obviously, this includes all Feynman diagrams in which a particle is
exchanged between the production subprocess and one of the decay
subprocesses or between the decay subprocesses. Examples for such
manifestly non-factorizable corrections are the diagrams (a), (b), and
(c) in \reffi{virtual_final_final_diagrams}.
\bfi
\begin{center}
\begin{picture}(360,230)(0,0)
\Text(0,220)[lb]{(a) type (\mfp)}
\Text(210,220)[lb]{(b) type (\mfp)}
\Text(0,90)[lb]{(c) type (\ffp)}
\Text(210,90)[lb]{(d) type (\mmp)}
\put(20,115){
\begin{picture}(150,100)(0,0)
\ArrowLine(30,50)( 5, 95)
\ArrowLine( 5, 5)(30, 50)
\Photon(30,50)(90,20){2}{6}
\Photon(30,50)(90,80){-2}{6}
\Vertex(60,65){2.0}
\GCirc(30,50){10}{.5}
\Vertex(90,80){2.0}
\Vertex(90,20){2.0}
\ArrowLine(90,80)(120, 95)
\ArrowLine(120,65)(90,80)
\ArrowLine(120, 5)( 90,20)
\ArrowLine( 90,20)(105,27.5)
\ArrowLine(105,27.5)(120,35)
\Vertex(105,27.5){2.0}
\Photon(60,65)(105,27.5){-2}{5}
\put(86,50){$\gamma$}
\put(63,78){$W$}
\put(40,65){$W$}
\put(52,18){$W$}
\put(10, 5){$\mathrm{e}^-(p_-)$}
\put(10,90){$\mathrm{e}^+(p_+)$}
\put(125,90){$f_1(k_1)$}
\put(125,65){$\bar f_2(k_2)$}
\put(125,30){$f_3(k_3)$}
\put(125, 5){$\bar f_4(k_4)$}
\end{picture}
}
\put(210,115){
\begin{picture}(120,100)(0,0)
\ArrowLine(30,50)( 5, 95)
\ArrowLine( 5, 5)(30, 50)
\Photon(30,50)(90,80){-2}{6}
\Photon(30,50)(90,20){2}{6}
\Vertex(60,35){2.0}
\GCirc(30,50){10}{.5}
\Vertex(90,80){2.0}
\Vertex(90,20){2.0}
\ArrowLine(90,80)(120, 95)
\ArrowLine(120,65)(105,72.5)
\ArrowLine(105,72.5)(90,80)
\Vertex(105,72.5){2.0}
\ArrowLine(120, 5)(90,20)
\ArrowLine(90,20)(120,35)
\Photon(60,35)(105,72.5){2}{5}
\put(86,46){$\gamma$}
\put(63,13){$W$}
\put(40,24){$W$}
\put(55,73){$W$}
\end{picture}
}
\put(20,-15){
\begin{picture}(120,100)(0,0)
\ArrowLine(30,50)( 5, 95)
\ArrowLine( 5, 5)(30, 50)
\Photon(30,50)(90,80){-2}{6}
\Photon(30,50)(90,20){2}{6}
\GCirc(30,50){10}{.5}
\Vertex(90,80){2.0}
\Vertex(90,20){2.0}
\ArrowLine(90,80)(120, 95)
\ArrowLine(120,65)(105,72.5)
\ArrowLine(105,72.5)(90,80)
\Vertex(105,72.5){2.0}
\ArrowLine(120, 5)( 90,20)
\ArrowLine( 90,20)(105,27.5)
\ArrowLine(105,27.5)(120,35)
\Vertex(105,27.5){2.0}
\Photon(105,27.5)(105,72.5){2}{4.5}
\put(93,47){$\gamma$}
\put(55,73){$W$}
\put(55,16){$W$}
\end{picture}
}
\put(210,-15){
\begin{picture}(120,100)(0,0)
\ArrowLine(30,50)( 5, 95)
\ArrowLine( 5, 5)(30, 50)
\Photon(30,50)(90,80){-2}{6}
\Photon(30,50)(90,20){2}{6}
\Photon(70,30)(70,70){2}{3.5}
\Vertex(70,30){2.0}
\Vertex(70,70){2.0}
\GCirc(30,50){10}{.5}
\Vertex(90,80){2.0}
\Vertex(90,20){2.0}
\ArrowLine(90,80)(120, 95)
\ArrowLine(120,65)(90,80)
\ArrowLine(120, 5)( 90,20)
\ArrowLine( 90,20)(120,35)
\put(76,47){$\gamma$}
\put(45,68){$W$}
\put(45,22){$W$}
\put(72,83){$W$}
\put(72,11){$W$}
\end{picture}
}
\end{picture}
\end{center}
\caption{Examples of non-factorizable photonic corrections
in ${\cal O}(\alpha)$. The shaded blobs stand for all tree-level graphs
contributing to $\Pep\Pem\to\PWp\PWm$.
Whenever Feynman diagrams with intermediate would-be Goldstone bosons
$\phi^\pm$ instead of $W^\pm$ bosons are relevant, the inclusion of such
graphs is implicitly understood.
}
\label{virtual_final_final_diagrams}
\efi
If the additional exchanged particle is massive, the corresponding 
correction has no double pole for on-shell 
\PW~bosons. However, if a photon is
exchanged between the different subprocesses, this leads to a
doubly-resonant contribution originating from the soft-photon region.
This can be directly seen from the usual soft-photon approximation
(SPA), which yields contributions proportional to the (doubly-resonant)
lowest-order contribution.

The doubly-resonant contributions can be extracted on the basis of a
simple power-counting argument.
For instance, the loop integral corresponding to diagram (c) in
\reffi{virtual_final_final_diagrams} is of the following form:
\beqar\label{vintexp} I&=&
\int\!\rd^4q
\frac{N(q,k_i)}{(q^2-\la^2)[(q-k_3)^2-m_3^2]
  [(q-k_-)^2-\MW^2][(q+k_+)^2-\MW^2][(q+k_2)^2-m_2^2]}\nl &=&
\int\!\rd^4q
\frac{N(q,k_i)}{(q^2-\la^2)(q^2-2qk_3)
  (q^2-2qk_- + k_-^2-\MW^2)(q^2+2qk_+ + k_+^2-\MW^2)(q^2+2qk_2)},
\nln
\eeqar
where we have introduced an infinitesimal photon mass $\la$ to regularize the
infrared (IR) singularity. The function $N(q,k_i)$ involves the
numerator of the Feynman integral, i.e.\ a polynomial in the momenta $q$ and
$k_i$, and possible further denominator factors originating from propagators 
(hidden in the blob of the diagrams)
that are regular for $q=0$ and $k_\pm^2=\MW^2$.  
For on-shell \PW~bosons ($k_\pm^2=\MW^2$), 
the integral has a quadratic IR singularity. For off-shell \PW~bosons,
part of the IR singularity is regularized by the
off-shellness, $k_\pm^2-\MW^2\ne 0$, such that the usual logarithmic IR
singularity remains. 
Vice versa, the off-shell result develops a pole
if either \PW~boson becomes on shell, and is thus doubly-resonant. 
Thus, the quadratic IR singularity in the on-shell limit is characteristic 
of the doubly-resonant non-factorizable contributions. All terms that
involve a factor $q$ in the numerator are less 
IR-singular and therefore do not lead to 
doubly-resonant contributions and
can be omitted. Similarly, $q$ can be neglected in all denominator
factors included in $N(q,k_i)$. 
In summary, $q$ can be put to zero in $N(q,k_i)$ in DPA. 
We have checked this for various examples explicitly.
As a consequence, we are left with only scalar integrals,
and the non-factorizable virtual corrections are proportional to the
lowest-order matrix element. We call the resulting approximation
{\it extended soft-photon approximation} (ESPA). It differs from the
usual SPA only by the fact that $q$ is not
neglected in the resonant \PW~propagators. In ESPA,
diagram (c) in \reffi{virtual_final_final_diagrams} gives the following
contribution to the matrix element:
\beqar\label{vintESPA} 
\M^{\bar f_2 f_3} &=& \ri e^2 Q_2Q_3\M_\born
\int\!\frac{\rd^4q}{(2\pi)^4}
\frac{4k_2k_3}{(q^2-\la^2)(q^2-2qk_3)(q^2+2qk_2)}
\nl &&
\phantom{e^2 Q_2Q_3\M_\born\int\!\frac{\rd^4q}{(2\pi)^4}} \times
\frac{(k_-^2-\MW^2)(k_+^2-\MW^2)}{[(q-k_-)^2-\MW^2][(q+k_+)^2-\MW^2]}.
\eeqar

The $q^2$ terms in the last four denominators are not relevant in
the soft-photon limit and were omitted in \citeres{Me96,Be97a,Be97b}. In fact,
using the above power-counting argument it can easily be seen that the
differences of doubly-resonant 
contributions with and without these $q^2$
terms are non-doubly-resonant. We have chosen to keep the $q^2$ terms,
because we want to use the standard techniques for the evaluation of 
virtual scalar integrals \cite{tH79}. In DPA, \ie if we perform the
limit $k_\pm^2\to\MW^2$ after evaluating the integral, we should
obtain the same result.

In order to arrive at physical results, we have to incorporate the finite
width of the \PW~bosons. In DPA this can be done in 
at least two different ways:

As a first possibility, we perform the integrals for zero width and
afterwards put $k_\pm^2=\MW^2$ where this does not give rise to
singularities. In all other places, \ie in the resonant propagators
and in logarithms of the form $\ln(k_\pm^2-\MW^2+\ieps)$,
we replace $k_\pm^2-\MW^2+\ieps$ by $K_\pm=k_\pm^2-\MW^2+\ri\MW\GW$.
Since the width is only relevant in the on-shell limit, it is clear
that the (physical) on-shell width has to be used.

Alternatively, we introduce the width in the \PW~propagators before
integration. This has to be done with caution. If we introduce the
finite width by resumming \PW-self-energy insertions, the width depends on
the invariant mass of the \PW~boson and thus on the integration
momentum. Fortunately, the contribution we are interested in results
only from the soft-photonic region where the virtual \PW~bosons are
almost on shell. Therefore, we can insert the 
on-shell width inside the loop integral. After
performing the integral, we put $k_\pm^2=\MW^2$ and $\GW=0$ where this
does not lead to singularities. In DPA this gives the same results as
the above treatment.

In the following we write $M^2=\MW^2-\ri\MW\GW$ instead of $\MW^2$ in
the loop integrals. It is always understood that
$M^2$ and $k_\pm^2$ are replaced by $\MW^2$ where possible
after evaluation of the integrals. 

If we implement the width into the integrand, it is clear that only
the part of the integration region with $|q_0|\lsim\GW$ contributes 
in DPA. If $|q_0|\gg\GW$, one of the \PW~propagators must be
non-resonant and the contribution becomes negligible.

Once the width is introduced, it becomes evident that the 
relative error of the DPA is of the order of
$\GW/\mathrm{scale}$. Let $E_\CM$ be the centre-of-mass (CM) energy and 
$\De E= E_\CM-2\MW$ be the available kinetic
energy of the \PW~bosons. Then, for $\De E\gsim\MW$ the scale is
given by \MW, for $\GW\lsim\De E\lsim\MW$ it is given by $\De E$ and
for $\De E\lsim\GW$ it is given by $\GW$. This shows that the DPA is
only sensible several $\GW$'s above threshold.
This is simply due to the fact that, close to threshold,
the phase space where both \PW~propagators can
become doubly-resonant is very small,
and the singly-resonant diagrams become important.

\subsection{Classification and gauge-independent definition of the 
non-factorizable doubly-reso\-nant virtual corrections}

Manifestly non-factorizable corrections arise from photon exchange
between the final states of the two \PW~bosons (\ffp),
between initial and final state (if), and between one of the intermediate
resonant \PW~bosons and the final state of the other \PW~boson (\mfp). 
Examples for these types of corrections are shown in 
\reffi{virtual_final_final_diagrams} (c),
\reffi{virtual_initial_final_diagrams} (a), and
\reffi{virtual_final_final_diagrams} (a,b), respectively.
\bfi
\begin{center}
\begin{picture}(360,230)(0,0)
\Text(0,220)[lb]{(a) type (if)}
\Text(210,220)[lb]{(b) type (im)}
\Text(0,90)[lb]{(c) type (mf)}
\Text(210,90)[lb]{(d) type (mm)}
\put(20,115){
\begin{picture}(120,100)(0,0)
\ArrowLine(27,55)(15, 75)
\Vertex(15,75){2.0}
\ArrowLine(15,75)( 3, 95)
\ArrowLine( 3, 5)(30, 50)
\Photon(30,50)(90,80){-2}{6}
\Photon(30,50)(90,20){2}{6}
\GCirc(30,50){10}{.5}
\Vertex(90,80){2.0}
\Vertex(90,20){2.0}
\ArrowLine(90,80)(105,87.5)
\ArrowLine(105,87.5)(120, 95)
\ArrowLine(120,65)(90,80)
\ArrowLine(120, 5)( 90,20)
\ArrowLine( 90,20)(120,35)
\Vertex(105,87.5){2.0}
\PhotonArc(66.25,36.25)(64.25,52.9,142.9){2}{8}
\put(55,90){$\gamma$}
\put(68,55){$W$}
\put(55,16){$W$}
\end{picture}
}
\put(230,115){
\begin{picture}(150,100)(0,0)
\ArrowLine(27,55)(15, 75)
\Vertex(15,75){2.0}
\ArrowLine(15,75)( 3, 95)
\ArrowLine( 3, 5)(30, 50)
\Photon(30,50)(90,20){2}{6}
\Photon(30,50)(90,80){-2}{6}
\Vertex(60,65){2.0}
\GCirc(30,50){10}{.5}
\Vertex(90,80){2.0}
\Vertex(90,20){2.0}
\ArrowLine(90,80)(120, 95)
\ArrowLine(120,65)(90,80)
\ArrowLine(120, 5)( 90,20)
\ArrowLine( 90,20)(120,35)
\PhotonArc(32.5,47.5)(32.596,32.47,122.47){2}{4.5}
\put(36,92){$\gamma$}
\put(75,61){$W$}
\put(51,48){$W$}
\put(52,18){$W$}
\end{picture}
}
\put(20,-15){
\begin{picture}(120,100)(0,0)
\ArrowLine(30,50)( 3, 95)
\ArrowLine( 3, 5)(30, 50)
\Photon(30,50)(90,80){-2}{6}
\Photon(30,50)(90,20){2}{6}
\Vertex(70,70){2.0}
\GCirc(30,50){10}{.5}
\Vertex(90,80){2.0}
\Vertex(90,20){2.0}
\ArrowLine(90,80)(105,87.5)
\Vertex(105,87.5){2.0}
\ArrowLine(105,87.5)(120, 95)
\ArrowLine(120,65)(90,80)
\ArrowLine(120, 5)(90,20)
\ArrowLine(90,20)(120,35)
\PhotonArc(87.5,78.75)(19.566,26.565,206.565){2}{6}
\put(57,86){$\gamma$}
\put(77,62){$W$}
\put(50,48){$W$}
\put(55,16){$W$}
\end{picture}
}
\put(230,-15){
\begin{picture}(120,100)(0,0)
\ArrowLine(30,50)( 5, 95)
\ArrowLine( 5, 5)(30, 50)
\Photon(30,50)(90,80){-2}{6}
\Photon(30,50)(90,20){2}{6}
\Vertex(75,72.5){2.0}
\Vertex(50,60){2.0}
\GCirc(30,50){10}{.5}
\Vertex(90,80){2.0}
\Vertex(90,20){2.0}
\ArrowLine(90,80)(120, 95)
\ArrowLine(120,65)(90,80)
\ArrowLine(120, 5)( 90,20)
\ArrowLine( 90,20)(120,35)
\PhotonArc(62.5,66.25)(13.975,26.565,206.565){-2}{3.5}
\put(55,90){$\gamma$}
\put(44,45){$W$}
\put(62,54){$W$}
\put(82,64){$W$}
\put(55,16){$W$}
\end{picture}
}
\end{picture}
\end{center}
\caption{Further examples of non-factorizable photonic corrections
in ${\cal O}(\alpha)$.}
\label{virtual_initial_final_diagrams}
\efi
In addition, there are diagrams where the photon 
does not couple to uniquely distinguishable subprocesses.
These contributions can be classified into photon-exchange 
contributions between one of the
intermediate resonant \PW~bosons and the final 
state of the same  
\PW~boson (mf), between the intermediate and the initial state (im), 
between the two intermediate \PW~bosons (\mmp), and within a single W-boson 
line, \ie the photonic part of the \PW-self-energy corrections (mm). 
Diagrams contributing to these types of corrections are given in 
\reffi{virtual_initial_final_diagrams} (c),
\reffi{virtual_initial_final_diagrams} (b),
\reffi{virtual_final_final_diagrams} (d), and
\reffi{virtual_initial_final_diagrams} (d), respectively.
Because the photon coupling to the \PW~boson can be
attributed to the decay or the production subprocesses, these diagrams
involve both factorizable and non-factorizable corrections.

In order to define the non-factorizable corrections, we have to specify
how the factorizable contributions are split off. This should be done
in such a way that the non-factorizable corrections become gauge-independent.
In \citeres{Be97a,Be97b} this was reached by exploiting the fact that 
in ESPA the matrix element can be viewed as a product of 
the lowest-order matrix element with two conserved currents. 
Taking all interferences between
the positively and the negatively charged currents 
arising from the outgoing W~bosons and fermions
gives a gauge-independent result.

We have chosen a different definition of the non-factorizable corrections,
which, however, turns out to be equivalent to the one of \citeres{Be97a,Be97b} 
in DPA. Our approach has the advantage of providing a clear procedure how 
to combine factorizable and non-factorizable contributions to the full
${\cal O}(\alpha)$ correction in DPA. Because the complete matrix
element is gauge-independent order by order, the sum of all
doubly-resonant $\Oa$ corrections must be gauge-independent. On the
other hand, the factorizable doubly-resonant corrections can be defined
by the product of gauge-independent on-shell matrix elements for 
W-pair production and \PW~decays and the 
(transverse parts of the) \PW~propagators, 
\beqar\label{mvirt}
{\cal M}_{\mathrm{f}} &=& 
\sum_{\la_+,\la_-} \frac{1}{K_+ K_-}
\Big( \de\M^\eeWW \M^\Wpff_\born \M^\Wmff_\born
\\ && {}
+ \M^\eeWW_\born \de\M^\Wpff \M^\Wmff_\born
+ \M^\eeWW_\born \M^\Wpff_\born \de\M^\Wmff \Big),
\nn
\eeqar
where $\de\M^\eeWW$, $\de\M^\Wpff$, and $\de\M^\Wmff$ denote the
one-loop amplitudes of the respective subprocesses. We can
define the non-factorizable doubly-resonant corrections by subtracting
the factorizable doubly-resonant corrections from the complete
doubly-resonant corrections. This definition allows 
us to calculate the complete doubly-resonant
corrections by simply adding the factorizable corrections, defined via
the on-shell matrix elements, to our results. Our definition
can be applied diagram by diagram. In this
way, all diagrams that are neither manifestly factorizable nor
manifestly non-factorizable can be split. 
Such diagrams receive doubly-resonant contributions from the complete
range of the photon momentum $q$, and not only from the soft-photon
region. This is obviously due to the presence of two explicit
resonant propagators. However, after subtracting the factorizable 
contributions, all 
doubly-resonant terms that are not IR-singular in the on-shell limit 
cancel exactly, i.e.\ only the soft-photon region contributes. 
Consequently, also in this case 
$q$ can be neglected everywhere except for the denominators that become
IR-singular in the on-shell limit.
As an example, we give the
non-factorizable correction originating from diagram (d) of
\reffi{virtual_final_final_diagrams}:%
\footnote{We use the sign $\sim$ to indicate an equality within 
DPA, \ie up to non-doubly-resonant terms.}
\beqar\label{virtual_coulomb_integral}
\M_{\mathrm{nf}}^{\PWp\PWm} & \sim & \ri e^2\M_\born\biggl\{
\int\!\frac{\rd^4q}{(2\pi)^4}\frac{4k_+k_-}
{q^2[(q + k_+)^2-M^2] [(q - k_-)^2-M^2]}\nl
&& \phantom{e^2\M_\born\biggl[}
-\biggl[\int\!\frac{\rd^4q}{(2\pi)^4}\frac{4k_+k_-}
{(q^2-\la^2)(q^2+2qk_+)(q^2-2qk_-)}
\biggr]_{k_\pm^2=\MW^2}\biggr\}.
\hspace{1em}
\eeqar
This example shows that the on-shell subtraction introduces
additional IR singularities. If the IR singularities in the 
non-factorizable real corrections are regularized in the same way, 
they cancel in the sum.
In \refeq{virtual_coulomb_integral} an infinitesimal photon mass
$\lambda$ is used as IR regulator, but we have repeated the same
calculation also by using a finite W-decay width as IR regulator instead
of $\lambda$, leading to the same results in the sum of virtual and
real photonic corrections.

We illustrate our definition of the non-factorizable corrections also for 
the photonic contribution to the \PW-self-energy correction 
[diagram (d) of \reffi{virtual_initial_final_diagrams}]. 
The non-factorizable part of the \PWp~self-energy reads 
\beqar\label{virtual_WSE_integral}
\M_{\mathrm{nf}}^{\PWp\PWp} & \sim & -\ri e^2\M_\born\biggl\{
\int\!\frac{\rd^4q}{(2\pi)^4}\frac{4k_+^2}{q^2[(q + k_+)^2-M^2](k_+^2-M^2)}\nl
&& \phantom{-\ri e^2\M_\born\biggl\{}
-\biggl[\int\!\frac{\rd^4q}{(2\pi)^4}\frac{4k_+^2}{q^2(q^2+2qk_+)}
\biggr]_{k_+^2=M^2} \frac{1}{k_+^2-M^2} \nl
&& \phantom{-\ri e^2\M_\born\biggl\{}
+\biggl[\int\!\frac{\rd^4q}{(2\pi)^4}\frac{4k_+^2}{(q^2-\la^2)(q^2+2qk_+)^2}
\biggr]_{k_+^2=\MW^2}\biggr\}.
\eeqar
The first integral results from the off-shell self-energy diagram, the second
from the corresponding mass-renormalization term, and the third
integral is the negative of the on-shell limit of the first two integrals.
The integrals in \refeq{virtual_WSE_integral} are UV-divergent and can
be easily evaluated in dimensional regularization. 

The gauge independence of the non-factorizable corrections has been
ensured by construction. The consistent evaluation of gauge theories
requires, besides gauge independence of the physical matrix elements,
the validity of Ward identities. 
It was found in \citere{bhf} that the violation of Ward identities can lead
to completely wrong predictions. The procedure described above
for extracting the non-factorizable corrections from the full matrix
element does not lead to problems with the Ward identities that rule the 
gauge cancellations inside matrix elements. This is due to the fact that
the non-factorizable corrections are proportional to the Born matrix 
element. Therefore, if Ward identities and gauge cancellations 
are under control in lowest order, the same is true for the non-factorizable 
corrections.

Finally, we show how our definition of the non-factorizable corrections
can be rephrased in terms of products of appropriately defined currents.
By using 
\beq \label{partfrac}
\frac{1}{(q\pm k_\pm)^2 - M^2} = \frac{1}{q^2\pm 2qk_\pm}
\biggl[1-\frac{k_\pm^2-M^2}{(q\pm k_\pm)^2 - M^2}\biggr],
\eeq
and the fact that in DPA
$k_\pm^2$ can be put to $\MW^2$ before integration
in integrals that do not depend on $M^2$,
the contribution \refeq{virtual_coulomb_integral} can be expressed as
\beqar\label{virtual_coulomb_integral2}
\M_{\mathrm{nf}}^{\PWp\PWm}&\sim& \ri e^2\M_\born
\int\!\frac{\rd^4q}{(2\pi)^4}\frac{4k_+k_-}{(q^2-\la^2)(q^2+2qk_+)(q^2-2qk_-)}
\biggl[-\frac{k_+^2-M^2}{(q+k_+)^2-M^2}
\nl&& \phantom{e^2\M_\born} {}
-\frac{k_-^2-M^2}{(q-k_-)^2-M^2}
+\frac{k_+^2-M^2}{(q+k_+)^2-M^2}\frac{k_-^2-M^2}{(q-k_-)^2-M^2}
\biggr].
\hspace{2em}
\eeqar 

The other non-factorizable corrections that involve photons coupled to
\PW~bosons can be rewritten in a similar way. Finally, all non-factorizable 
virtual corrections can be cast into the following form:
\beqar\label{virtual_nonfac}
\M_{\mathrm{nf}}^{\virt} &\sim& \ri\M_\born \int\!\frac{\rd^4q}{(2\pi)^4}
\frac{1}{q^2-\la^2}\left[j^{\eeWW,\mu}_\virt j^\Wpff_{\virt,\mu}
+j^{\eeWW,\mu}_\virt j^\Wmff_{\virt,\mu}
\right. \nl &&
\phantom{e^2\M_\born \int\!\frac{\rd^4q}{(2\pi)^4}\frac{1}{q^2-\la^2}\Big[}
\left.{}
+j^{\Wpff,\mu}_\virt j_{\virt,\mu}^\Wmff\right]
\eeqar
with
\beqar
\label{eq:virtcurr}
j_{\virt,\mu}^\eeWW &=&
e\biggl( \frac{2k_{+\mu}}{q^2+2q k_+} + \frac{2k_{-\mu}}{q^2-2q k_-}
      - \frac{2p_{-\mu}}{q^2-2q p_-} - \frac{2p_{+\mu}}{q^2+2q p_+}\biggr)\nlc
j_{\virt,\mu}^\Wpff&=&
e\biggl( Q_1 \frac{2k_{1\mu}}{q^2+2q k_1} - Q_2 \frac{2k_{2\mu}}{q^2+2q k_2}
      - \frac{2k_{+\mu}}{q^2+2q k_+}\biggr)\frac{k_+^2-M^2}{(k_++q)^2-M^2} \nlc
j_{\virt,\mu}^\Wmff&=&
-e\biggl( Q_3 \frac{2k_{3\mu}}{q^2-2q k_3} - Q_4 \frac{2k_{4\mu}}{q^2-2q k_4}
      + \frac{2k_{-\mu}}{q^2-2q k_-}\biggr)\frac{k_-^2-M^2}{(k_--q)^2-M^2}.
\hspace{2em}
\eeqar
The last term in \refeq{virtual_nonfac} originates from the Feynman graphs 
shown in \reffi{virtual_final_final_diagrams} and 
those where the final-state fermions are appropriately interchanged 
[interference terms (\ffp), (\mfp), and (\mmp)].
The contributions involving the current $j_{\virt,\mu}^\eeWW$ contain 
the interference terms (if), (mf), (im), (mm), and the remaining
contributions of (\mfp) and (\mmp).
The contribution of the \PWp~self-energy is given, for instance, 
by the product of the two terms involving $k_{+\mu}$ in 
$j^{\eeWW}_{\virt,\mu}$ and $j_{\virt,\mu}^\Wpff$.

In DPA, the $q^2$ terms in the denominators 
of \refeq{eq:virtcurr} can be neglected, and the
currents are conserved. The currents $j_{\virt,\mu}^\Wpff$
and $j_{\virt,\mu}^\Wmff$ are the ones mentioned in \citeres{Be97a,Be97b}.
For the virtual corrections,
this shows that our definition of non-factorizable doubly-resonant corrections
coincides with the one of \citeres{Be97a,Be97b} in DPA.
 
\subsection{Doubly-resonant real corrections}

The photonic virtual corrections discussed above are IR-singular and
have to be combined with the corresponding real corrections in order
to arrive at a sensible physical result.
The real corrections originate from the process
\beqar\label{eeffffa}
\Pep(p_+) + \Pem(p_-) & \;\to\; & \PWp(k'_+) + \PWm(k'_-)\, [{}+ \ga(q)]
\nl
& \;\to\; &
f_1(k'_1) + \bar f_2(k'_2) + f_3(k'_3) + \bar f_4(k'_4)+ \ga(q). 
\eeqar
Note that we have marked the fermion momenta $k'_i$ by primes in order
to distinguish them from the respective momenta without real photon
emission. 
The momenta of the W~bosons are $k'_+=k'_1+k'_2$ and $k'_-=k'_3+k'_4$
if the photon is emitted in the initial state or in the final state of
the other \PW~boson, and $\bar k_+=k'_1+k'_2+q$ or $\bar k_-=k'_3+k'_4+q$
if the photon is emitted in the final state of the respective \PW~boson.

The non-factorizable corrections 
induced by the process \refeq{eeffffa} arise 
from interferences between diagrams where the photon is
emitted from different subprocesses.
A typical non-factorizable contribution is shown in
\reffi{real_non_factorizable_diagram}. 
\bfi
\centerline{
\begin{picture}(240,100)(0,0)
\ArrowLine(30,50)( 5, 95)
\ArrowLine( 5, 5)(30, 50)
\Photon(30,50)(90,80){-2}{6}
\Photon(30,50)(90,20){2}{6}
\GCirc(30,50){10}{.5}
\Vertex(90,80){2.0}
\Vertex(90,20){2.0}
\ArrowLine(90,80)(120, 95)
\ArrowLine(120,65)(105,72.5)
\ArrowLine(105,72.5)(90,80)
\ArrowLine(120, 5)( 90,20)
\ArrowLine( 90,20)(120,35)
\Vertex(105,72.5){2.0}
\PhotonArc(120,65)(15,150,270){2}{3}
\put(55,73){$W$}
\put(55,16){$W$}
\put(100,47){$\gamma$}
\DashLine(120,0)(120,100){6}
\PhotonArc(120,35)(15,-30,90){2}{3}
\Vertex(135,27.5){2.0}
\ArrowLine(150,80)(120,95)
\ArrowLine(120,65)(150,80)
\ArrowLine(120, 5)(150,20)
\ArrowLine(150,20)(135,27.5)
\ArrowLine(135,27.5)(120,35)
\Vertex(150,80){2.0}
\Vertex(150,20){2.0}
\Photon(210,50)(150,80){2}{6}
\Photon(210,50)(150,20){-2}{6}
\ArrowLine(210,50)(235,95)
\ArrowLine(235, 5)(210,50)
\GCirc(210,50){10}{.5}
\put(177,73){$W$}
\put(177,16){$W$}
\end{picture}
}
\caption{Example of a non-factorizable real correction.}
\label{real_non_factorizable_diagram}
\efi
Including the integration over the
photon phase space, this contribution 
has the following form:
\beqar\label{rintex}
\Ibr&=&
\int\!\frac{\rd^3\bq}{2q_0}
\frac{N(q,k'_i)\de(p_++p_--k'_1-k'_2-k'_3-k'_4-q)}
{[(q+k'_2)^2-m_2^2][(q+k'_+)^2-M^2](k_-^{\prime 2}-M^2)}
\nn\\*
&& {} \phantom{\int\!\frac{\rd^3\bq}{2q_0}} \times
\Biggl\{ \frac{1}{[(q+k'_3)^2-m_3^2](k_+^{\prime 2}-M^2)[(q+k'_-)^2-M^2]}
\Biggr\}^*
\,\biggr|_{q_0=\sqrt{{\bf q}^2+\la^2}}
\nl
&=&\int\!\frac{\rd^3\bq}{2q_0}
\frac{N(q,k'_i)\de(p_++p_--k'_1-k'_2-k'_3-k'_4-q)}
{(2qk'_2)(2qk'_+ + k_+^{\prime 2}-M^2)(k_-^{\prime 2}-M^2)}
\nn\\*
&& {} \phantom{\int\!\frac{\rd^3\bq}{2q_0}} \times
\frac{1}{(2qk'_3)[k_+^{\prime 2}-(M^*)^2][2qk'_- + k_-^{\prime 2}-(M^*)^2]}
\,\biggr|_{q_0=\sqrt{{\bf q}^2+\la^2}},
\eeqar
where we again use a photon mass $\la$ to regularize the IR
singularities, and $N(q,k'_i)$ has the same meaning as above.
Again, the doubly-resonant contributions are characterized by a
quadratic IR singularity for $k_\pm^{\prime 2}=\MW^2$, 
$\Gamma_\PW\to 0$, and only soft-photon emission is relevant in DPA. 
For this reason, the \PW~bosons are nearly on shell, and the on-shell width
is appropriate. As for the virtual corrections, the introduction
of the width before or after phase-space integration leads to the same
results in DPA. As already indicated in \refeq{rintex},
in the following real integrals we use $M^2$ 
with the understanding that it has to be replaced by $\MW^2$ after
integration where possible.

The aim is to integrate over the photon momentum analytically and to 
relate the fermion momenta $k'_i$ to the ones of the process without photon 
emission, $k_i$. Primed and unprimed momenta differ
by terms of the order of the photon momentum: $k_i'=k_i+\O(q_0)$.
In DPA we can neglect $q$ in $N(q,k'_i)$,
leading to the replacement $N(q,k'_i)\to N(0,k_i)$.
Moreover, we can extend the integration region for $q_0$ to infinity,
because large photon momenta yield negligible contributions in DPA. 
After extension of the integration region the integral becomes
Lorentz-invariant. 

While the correction factor to the lowest-order cross section 
is universal in SPA for all
observables, the correction factor is non-universal in ESPA. In order to
define this correction factor in a unique way, one has to specify the
parametrization of phase space, i.e.\ the variables that are kept fixed
when the photon momentum is integrated over. 
This fact has not been addressed in the literature so far. 

Let us consider this problem in more detail.
It can be traced back to the
appearance of the photon momentum $q$ in the $\de$-function for momentum
conservation. In the usual SPA $q$ is neglected in this $\de$-function, 
which is sensible if the exact matrix element is a slowly varying function
of $q$ in the vicinity of $q=0$. However, in the presence of resonant
propagators, in which $q$ cannot be neglected, the simple omission of $q$
in the  momentum-conservation $\de$-function leads to ambiguous results:
putting $q=0$ in the $\de$-function
and identifying $k'_i$ with $k_i$ in \refeq{rintex} yields
\beqar\label{rintex1}
\Ibr &\to&
\int\!\frac{\rd^3\bq}{2q_0}
\frac{N(0,k_i)\de(p_++p_--k_1-k_2-k_3-k_4)}
{(2qk_2)(2qk_+ + k_+^2-M^2)(k_-^2-M^2)}
\nn\\*
&& {} \phantom{\int\!\frac{\rd^3\bq}{2q_0}} \times
\frac{1}{(2qk_3)[k_+^2-(M^*)^2][2qk_- + k_-^2-(M^*)^2]}
\,\biggr|_{q_0=\sqrt{{\bf q}^2+\la^2}}.
\eeqar 
On the other hand, eliminating $k'_+$ in the denominator of
\refeq{rintex} with the help of the $\de$-function, putting $q=0$ in
the $\de$-function, restoring 
$k'_+$ with the modified $\de$-function, and setting $k'_i\to k_i$ results in
\beqar\label{rintex2}
\Ibr &\to&
\int\!\frac{\rd^3\bq}{2q_0}
\frac{N(0,k_i)\de(p_++p_--k_1-k_2-k_3-k_4)}
{(2qk_2)(k_+^2-M^2)(k_-^2-M^2)}
\nn\\*
&& {} \phantom{\int\!\frac{\rd^3\bq}{2q_0}} \times
\frac{1}{(2qk_3)[-2qk_+ + k_+^2-(M^*)^2][2qk_- + k_-^2-(M^*)^2]}
\,\biggr|_{q_0=\sqrt{{\bf q}^2+\la^2}}.
\eeqar
Both expressions differ by a doubly-resonant contribution. The
difference is in general confined to the \PW~propagators and
originates from the fact that not only soft photons but also photons
with energies of the order of $|k_\pm^2-\MW^2|/\MW$ or, after the
inclusion of the finite width, of order $\GW$ contribute in DPA.
Since photons with finite energies contribute, it is evident
that the integral over the photon momentum depends on the choice of
the phase-space variables that are kept fixed.

As a consequence, one has to choose a definite
parametrization of phase space and to exploit the $\de$-function
carefully, in order to define the non-factorizable corrections uniquely.
For instance, if the vector
$k'_+=k'_1+k'_2$ is kept fixed, the alternative \refeq{rintex2} is
excluded. However, because of momentum conservation,
not all external momenta can be kept fixed independently. 

It is, however, possible to keep, for instance, the invariant masses of the
final-state fermion pairs $k_+^{\prime2}=(k'_1+k'_2)^2$ and
$k_-^{\prime2}=(k'_3+k'_4)^2$ fixed  when integrating over the photon momentum.
If we require $(k'_1+k'_2)^2=k_+^{\prime2}=k_+^2=(k_1+k_2)^2$, we
obtain for the denominator of the \PWp~boson
\beqar\label{Wpprop}
(q+k'_+)^2-M^2 &=& 2qk'_++ k_+^{\prime2} - M^2  
= 2qk_++ k_+^{2} - M^2 + \O(q_0^2) \nl
&=& (q+k_+)^2-M^2 + \O(q_0^2),
\eeqar
where $k_i'=k_i+\O(q_0)$ was used.
Based on the power-counting argument given above,
the terms of order $q_0^2$ can be neglected in DPA, and we find
\beq
(q+k'_+)^2-M^2 \sim (q+k_+)^2-M^2 .
\eeq
If we choose to eliminate $k'_+$, as done in the derivation of 
\refeq{rintex2}, we find, on the other hand,
\beqar
(q+k'_+)^2-M^2 &=& (p_++p_--k'_-)^2-M^2 
=(p_++p_--k_-)^2-M^2 + \O(q_0) \nl
&=&  k_+^2 - M^2 + \O(q_0).
\eeqar
The $\O(q_0)$ terms are relevant in DPA [and in fact given by \refeq{Wpprop}]. 
As a consequence, \refeq{rintex2} is not correct if
we choose to fix $k_+^{\prime2}=k_+^2$ when integrating over the
photon momentum. For fixed $k_+^{\prime2}=k_+^2$, \refeq{Wpprop} leads to the
unique result \refeq{rintex1} for the \PWp~propagator in 
DPA, independently of the other phase-space parameters. 
If we choose, on the other hand, to fix $\bar k_+^2=k_+^2$,
which corresponds to a different 
definition of the invariant mass of the \PWp~boson, we obtain
\beq
(q+k'_+)^2-M^2 = \bar k_+^2 - M^2 ,
\eeq
and thus \refeq{rintex2} instead of \refeq{rintex1}. 
Consequently, the different approaches \refeq{rintex1} and
\refeq{rintex2} correspond to different definitions of the invariant
mass of the \PWp~boson which decays into the fermion pair $(f_1,\bar f_2)$. 
In order to
define the DPA for real radiation, one has to specify at least the
definition of the invariant masses of the \PW~bosons that are kept fixed.
In the following we always fix $k_+^2=(k'_1+k'_2)^2=(k_1+k_2)^2$
and $k_-^2=(k'_3+k'_4)^2=(k_3+k_4)^2$,
as it was also implicitly done in \citeres{Me96,Be97a,Be97b}. 
Once the invariant masses of the \PW~bosons are fixed in this way, the
resulting formulae for the non-factorizable corrections hold
independently of the choice of all other phase-space variables.

We stress that the results obtained within this parametrization of 
phase space differ from those in other parametrizations by
doubly-resonant corrections. 
As already indicated in the introduction, in an experimentally more
realistic approach the invariant masses of W~bosons are identified with
invariant masses of jet pairs, which also include part of the
photon radiation. Since this situation can only be described with 
Monte Carlo programs, 
our results (as well as those of \citeres{Me96,Be97a,Be97b}) should be 
regarded as an estimate of the non-factorizable corrections. 
Moreover, our analytic results provide benchmarks for such future 
Monte Carlo calculations.  

\subsection{Classification and gauge-independent definition of the 
non-factorizable doubly-resonant real corrections}

The doubly-resonant real corrections can be classified in exactly the same 
way as the virtual corrections. For each virtual diagram there is exactly 
a real contribution, which we denote in the same way, e.g.\ \ffp\
refers to all interferences where the photon is emitted by two fermions
corresponding to the two different W~bosons. 

As in the case of the virtual corrections, one has to define the
non-factorizable real corrections in a gauge-independent way. 
To this end, we proceed analogously and 
define the non-factorizable real corrections as the difference of the
complete real corrections and the factorizable real corrections in DPA. 
The factorizable corrections are defined by the products of the matrix
elements for on-shell W-pair production and decay with additional photon
emission. There are three contributions to the factorizable real corrections, 
one where the photon is radiated during the production process and two
where it is emitted during one of the \PW-boson decays. The corresponding 
matrix elements read 
\beqar\label{realfac}
{\cal M}_{\real,1} &=& 
\sum_{\la_+,\la_-}\frac{\M^{\eeWW\ga}_\born\M^\Wpff_\born \M^\Wmff_\born}
{(k_+^2-M^2)(k_-^2-M^2)}\nlc
{\cal M}_{\real,2} &=& 
\sum_{\la_+,\la_-}\frac{\M^\eeWW_\born \M^{\Wpff\ga}_\born\M^\Wmff_\born}
{[(k_++q)^2-M^2](k_-^2-M^2)}\nlc
{\cal M}_{\real,3} &=& 
\sum_{\la_+,\la_-}\frac{\M^\eeWW_\born \M^\Wpff_\born \M^{\Wmff\ga}_\born}
{(k_+^2-M^2)[(k_-+q)^2-M^2]},
\eeqar
in analogy to \refeq{mborn}.
Note that the matrix elements $\M_{\real,2}$ and $\M_{\real,3}$
involve an explicit $q$-dependent propagator.
The factorizable corrections are given by
the squares of these three matrix elements, and 
include by definition no interferences between them.

As an example for the extraction of non-factorizable corrections from real 
diagrams that involve both factorizable and non-factorizable corrections,
we consider the contribution of the diagram in \reffi{real_mixed_diagram}.
\bfi
\centerline{
\begin{picture}(240,100)(0,0)
\ArrowLine(30,50)( 5, 95)
\ArrowLine( 5, 5)(30, 50)
\Photon(30,50)(90,80){-2}{6}
\Photon(30,50)(90,20){2}{6}
\GCirc(30,50){10}{.5}
\Vertex(90,80){2.0}
\Vertex(90,20){2.0}
\ArrowLine(90,80)(120, 95)
\ArrowLine(120,65)(90,80)
\ArrowLine(120, 5)( 90,20)
\ArrowLine( 90,20)(120,35)
\Vertex(65,67.5){2.0}
\put(55,73){$W$}
\put(55,16){$W$}
\DashLine(120,0)(120,100){6}
\Photon(65,67.5)(175,32.5){-2}{10}
\put(100,45){$\gamma$}
\Vertex(175,32.5){2.0}
\ArrowLine(150,80)(120,95)
\ArrowLine(120,65)(150,80)
\ArrowLine(120, 5)(150,20)
\ArrowLine(150,20)(120,35)
\Vertex(150,80){2.0}
\Vertex(150,20){2.0}
\Photon(210,50)(150,80){2}{6}
\Photon(210,50)(150,20){-2}{6}
\ArrowLine(210,50)(235,95)
\ArrowLine(235, 5)(210,50)
\GCirc(210,50){10}{.5}
\put(177,73){$W$}
\put(177,16){$W$}
\end{picture}
}
\caption{Real bremsstrahlung diagram containing non-factorizable and factorizable contributions.}
\label{real_mixed_diagram}
\efi
After subtraction of the factorizable contribution, which 
originates from $|\M_{\real,1}|^2$, it gives rise to
the following correction factor to the 
square $|\M_\born|^2$ of the lowest-order matrix element \refeq{mborn}:
\beqar\label{rWWnfcorr}
\de_{\real}^{\PWp\PWm} &=&e^2 \int\!\frac{\rd^3\bq}{(2\pi)^3 2q_0}\, 
2\Re\biggl\{
\frac{4k_+k_-}{[(k_++q)^2-M^2][(k_-+q)^2-(M^*)^2]}
\nn\\ && \qquad {}
-\biggl[\frac{4 k_+k_-}{2q k_+ 2q k_-}\biggr]_{k_\pm^2=\MW^2}\biggr\}
\bigg|_{q_0=\sqrt{{\bf q}^2+\la^2}}.
\eeqar
Note that the form of the correction factor is
only correct for fixed $(k'_1+k'_2)^2$ and $(k'_3+k'_4)^2$.
For other conventions the off-shell contribution changes,
whereas the on-shell contribution stays the same.
Using the relations \refeq{partfrac} for $q^2=0$,
in DPA we can rewrite \refeq{rWWnfcorr} as
\beqar
\de_{\real}^{\PWp\PWm}
&\sim&-e^2\int\!\frac{\rd^3\bq}{(2\pi)^3 2q_0}\,
2\Re\biggl\{\frac{k_+k_-}{(q k_+)(q k_-)} \biggl[ 
           \frac{k_+^2-M^2}{(k_++q)^2-M^2}
         + \frac{k_-^2-(M^*)^2}{(k_-+q)^2-(M^*)^2}  
\nn\\ && \qquad {}
-\frac{k_+^2-M^2}{(k_++q)^2-M^2} \frac{k_-^2-(M^*)^2}{(k_-+q)^2-(M^*)^2}
\biggr]\biggr\}
\biggr|_{q_0=\sqrt{{\bf q}^2+\la^2}}. 
\eeqar

In the same way, all other contributions  that originate
from a photon coupled to a \PW~boson can be rewritten such that the
complete real non-factorizable corrections can finally be 
expressed as the following correction factor to the lowest-order cross section,
\beqar\label{rnfcorr}
\de_{\real,\nf} &\sim& -\int\!\frac{\rd^3\bq}{(2\pi)^3 2q_0}\,
2\Re\left[j^{\eeWW,\mu}_{\real} (j^\Wpff_{\real,\mu})^*
+j^{\eeWW,\mu}_{\real}(j^\Wmff_{\real,\mu})^*\right. \nl
&& \phantom{-\int\!\frac{\rd^3\bq}{(2\pi)^3 2q_0}\,2\Re\Big[}
\left.\left.{}+j^{\Wpff,\mu}_{\real}(j^\Wmff_{\real,\mu})^*\right]
\right|_{q_0=\sqrt{{\bf q}^2+\la^2}}
\eeqar
with the currents
\beqar
\label{realcurrents}
j_{\real,\mu}^\eeWW &=&
e\biggl( \frac{k_{+\mu}}{q k_+} - \frac{k_{-\mu}}{q k_-}
      + \frac{p_{-\mu}}{q p_-} - \frac{p_{+\mu}}{q p_+}\biggr)\nlc
j_{\real,\mu}^\Wpff&=&
e\biggl( Q_1 \frac{k_{1\mu}}{q k_1} - Q_2 \frac{k_{2\mu}}{q k_2}
      - \frac{k_{+\mu}}{q k_+}\biggr)\frac{k_+^2-M^2}{(k_++q)^2-M^2} ,\nl
j_{\real,\mu}^\Wmff&=&
e\biggl( Q_3 \frac{k_{3\mu}}{q k_3} - Q_4 \frac{k_{4\mu}}{q k_4}
      + \frac{k_{-\mu}}{q k_-}\biggr)\frac{k_-^2-M^2}{(k_-+q)^2-M^2}.
\eeqar  

The factor \refeq{rnfcorr} for the non-factorizable correction
can be viewed as the interference contributions in the square 
of the matrix element ($\veps$ denotes the polarization vector of the photon) 
\beq
{\cal M}_{\real} =  {\cal M}_{\born}\,
\veps^\mu \left(j_{\real,\mu}^\eeWW+j_{\real,\mu}^\Wpff+j_{\real,\mu}^\Wmff\right),
\eeq
which is just the sum of the three matrix elements $\M_{\real,i}$,
$i=1,2,3$, in ESPA (including radiation from the external fermions
and the internal \PW~bosons). The respective squares
of these three contributions correspond to the factorizable corrections.  
Note that $\M_\born^\eeWW\*\veps^\mu j_{\real,\mu}^\eeWW$ is the
soft-photon matrix element for on-shell \PW-pair production.
Similarly, $\M_\born^\Wpff\*\veps^\mu j_{\real,\mu}^\Wpff$ and
$\M_\born^\Wmff\*\veps^\mu j_{\real,\mu}^\Wmff$ 
correspond to the soft-photon matrix elements for the
decays of the on-shell \PW~bosons, apart from the extra factors
$(k_\pm^2-M^2)/[(k_\pm+q)^2-M^2]$. These factors result from the
definition of the lowest-order matrix element in terms of $k_\pm$ and
from the fact that we 
impose $k_\pm^{\prime2}=k_\pm^2$ and not $\bar k_\pm^2=k_\pm^2$.

Obviously, the currents \refeq{realcurrents} are conserved, and the
corresponding Ward identity for the U(1)$_{\mathrm{em}}$ symmetry of
the emitted photon is fulfilled.

\section{Calculation of scalar integrals}
\label{secalc}

In this section we set our conventions for the scalar integrals.
We describe the reduction of the virtual and real 5-point functions to
4-point functions and indicate how the scalar integrals were evaluated.
More details and the explicit results for the scalar integrals can be 
found in the appendix.

\subsection{Reduction of 5-point functions}

\subsubsection{Reduction of virtual 5-point functions}

The virtual 3-, 4-, and 5-point functions are defined as
\beqar \label{CDE0int}
C_0(p_1,p_2,m_0,m_1,m_2) &=&
\frac{1}{\ri\pi^{2}}\int\!\rd^{4}q\,\frac{1}{N_0 N_1 N_2},
\nl
D_{\{0,\mu\}}(p_1,p_2,p_3,m_0,m_1,m_2,m_3) &=&
\frac{1}{\ri\pi^{2}}\int\!\rd^{4}q\,\frac{\{1,q_\mu\}}{N_0 N_1 N_2 N_3},
\nl
E_{\{0,\mu\}}(p_1,p_2,p_3,p_4,m_{0},m_1,m_2,m_3,m_4) &=&
\frac{1}{\ri\pi^{2}}\int\!\rd^{4}q\,\frac{\{1,q_\mu\}}{N_0 N_1 N_2 N_3 N_4},
\eeqar
with the denominator factors
\beq \label{D0Di}
N_{0}= q^{2}-m_{0}^{2}+\ri\epsilon, \qquad
N_{i}= (q+p_{i})^{2}-m_{i}^{2}+\ri\epsilon, \qquad i=1,\ldots,4 ,
\eeq
where $\ri\epsilon$ $(\epsilon>0)$ 
denotes an infinitesimal imaginary part. 

The reduction of the virtual 5-point function to 4-point functions is
based on the fact that in four dimensions the integration momentum
depends linearly on the four external momenta $p_i$ \cite{Me65}.
This gives rise to the identity
\beq\label{detid}
0=\left\vert
\barr{cccc}
2q^{2}  & 2qp_{1}    & \ldots & \;2qp_{4} \\
2p_{1}q & 2p_{1}^{2} & \ldots & \;2p_{1}p_{4} \\
\vdots    & \vdots     & \ddots     &\;\vdots     \\
2p_{4}q & 2p_{4}p_{1} & \ldots &\; 2p_{4}^{2}
\earr
\right\vert =
\left\vert\barr{cccc}
2N_{0}+Y_{00}  & 2qp_{1}    & \ldots & \;2qp_{4} \\
N_{1}-N_{0}+Y_{10}-Y_{00} & 2p_{1}^{2} & \ldots & \;2p_{1}p_{4} \\
\vdots    & \vdots     & \ddots     &\;\vdots     \\
N_{4}-N_{0}+Y_{40}-Y_{00} & 2p_{4}p_{1} & \ldots &\; 2p_{4}^{2}
\earr\right\vert
\eeq
with
\beq\label{defY}
Y_{00} = 2 m_0^2, \quad Y_{i0} = Y_{0i} = m_0^2 + m_i^2-p_i^2, \quad
Y_{ij} = m_i^2 + m_j^2 - (p_i-p_j)^2, \quad i,j=1,2,3,4.
\eeq
Dividing this equation by $N_{0}N_{1}\cdots N_{4}$ and integrating over
$\rd^{4}q$ yields
\beq\label{E0red0}
0 = \frac{1}{\ri\pi^{2}}\int\! \rd^{4}q\, \frac{1}{N_{0}N_{1}\cdots N_{4}}
\left\vert
\barr{cccc}
2N_{0}+Y_{00}  & 2qp_{1}    & \ldots & \;2qp_{4} \\
N_{1}-N_{0}+Y_{10}-Y_{00} & 2p_{1}^{2} & \ldots & \;2p_{1}p_{4} \\
\vdots    & \vdots     & \ddots     &\;\vdots     \\
N_{4}-N_{0}+Y_{40}-Y_{00} & 2p_{4}p_{1} & \ldots &\; 2p_{4}^{2}
\earr
\right\vert.
\eeq
Expanding the determinant along the first column, we obtain
\beqar \label{E0red1}
0&=&\Bigl[2D_{0}(0)+Y_{00}E_0\Bigr]
\left\vert\barr{ccc}
 2p_{1}p_{1}    & \ldots & \;2p_{1}p_{4} \\
 \vdots     & \ddots     &\;\vdots     \\
 2p_{4}p_{1} & \ldots &\; 2p_{4}p_{4}
\earr\right\vert \nl
&&+\disp\sum_{k=1}^{4}(-1)^{k}\Bigl\{D_{\mu}(k) 
-\Bigl[D_{\mu}(0)+p_{4\mu}D_{0}(0)\Bigr]
+p_{4\mu}D_{0}(0)+(Y_{k0}-Y_{00})E_{\mu}\Bigr\}\nl
&&\qquad\qquad\times \left\vert\barr{ccc}
 2p_1^\mu    & \ldots & \;2p_4^\mu \\
 2p_1p_1 & \ldots &\; 2p_1p_4\\
 \vdots     & \ddots     &\;\vdots     \\
 2p_{k-1}p_1 & \ldots &\; 2p_{k-1}p_4\\
 2p_{k+1}p_1 & \ldots &\; 2p_{k+1}p_4\\
 \vdots     & \ddots     &\;\vdots     \\
 2p_4p_1 & \ldots &\; 2p_4p_4
\earr\right\vert,
\eeqar
where $D_0(k)$ denotes the 4-point function that is obtained from the
5-point function $E_0$ by omitting the $k$th propagator $N_k^{-1}$.
The terms involving $p_{4\mu}D_{0}(0)$ have been added for later convenience. 
 
All integrals in \refeq{E0red1} are UV-finite and
Lorentz-covariant. Therefore, the vector integrals possess the
following decompositions
\beqar\label{decomp}
E_\mu &=& \sum_{i=1}^{4} E_i p_{i\mu}, \nl 
D_\mu(k) &=& \sum_{i=1\atop  i\ne k }^4 D_i(k) p_{i\mu}, \qquad
k=1,2,3,4 \nlc
D_\mu(0)+p_{4\mu}D_0(0) &=& \sum_{i=1}^{3} D_i(0) (p_{i}-p_4)_\mu .
\eeqar
The last decomposition becomes obvious after performing a shift
$q\to q-p_4$ in the integral. From \refeq{decomp} it follows
immediately that the terms 
in \refeq{E0red1} that involve $D_\mu(k)$ drop out 
when multiplied with the determinants, because the resulting determinants
vanish. Similarly, $D_\mu(0)+p_{4\mu}D_0(0)$
vanishes after summation over $k$.  Finally, the term
$p_{4\mu}D_0(0)$ contributes only for $k=4$, where it
can be combined with the first term in (\ref{E0red1}).  Rewriting the
resulting equation as a determinant and reinserting the explicit form
of the tensor integrals leads to
\beq \label{eq46}
0 = \frac{1}{\ri\pi^{2}}\int\!\rd^{4}q\,
\frac{1}{N_{0}N_{1}\cdots N_{4}}
\left\vert
\barr{cccc}
N_{0}+Y_{00}  & 2qp_{1}    & \ldots & \;2qp_{4} \\
Y_{10}-Y_{00} & 2p_{1}p_1 & \ldots & \;2p_{1}p_{4} \\
\vdots    & \vdots     & \ddots     &\;\vdots     \\
Y_{40}-Y_{00} & 2p_{4}p_{1} & \ldots &\; 2p_{4}p_4
\earr
\right\vert.
\eeq
Using
\beqar
2p_{i}p_{j}&=& Y_{ij}-Y_{i0}-Y_{0j}+Y_{00},\nl
2qp_{j}&=& N_{j}-N_{0}+Y_{0j}-Y_{00} ,
\eeqar
adding the first column to each of the other columns, and then
enlarging the determinant by one column and one row, this can be written as
\beq\label{E0red2}
0 = \left\vert\barr{cccc}
1 & Y_{00}    & \ldots & \;Y_{04} \\
0 & \quad D_{0}(0)+Y_{00}E_{0} \quad & \ldots & \quad D_{0}(4)+Y_{04}E_{0} \\
0 & Y_{10}-Y_{00} & \ldots &\; Y_{14}-Y_{04}  \\
\vdots    & \vdots     & \ddots     &\;\vdots     \\
0 & Y_{40}-Y_{00} & \ldots &\; Y_{44}-Y_{04}
\earr\right\vert.
\eeq
Equation \refeq{E0red2} is equivalent to
\beq \label{E0red}
0 = \left\vert \barr{cccccc}
-E_{0} &\:D_{0}(0)&\:D_{0}(1)&\:D_{0}(2)&\:D_{0}(3)&\:D_{0}(4)\\
  1   &  Y_{00}   &  Y_{01}   &  Y_{02}   &  Y_{03}   &  Y_{04}   \\
  1   &  Y_{10}   &  Y_{11}   &  Y_{12}   &  Y_{13}   &  Y_{14}   \\
  1   &  Y_{20}   &  Y_{21}   &  Y_{22}   &  Y_{23}   &  Y_{24}   \\
  1   &  Y_{30}   &  Y_{31}   &  Y_{32}   &  Y_{33}   &  Y_{34}   \\
  1   &  Y_{40}   &  Y_{41}   &  Y_{42}   &  Y_{43}   &  Y_{44}
\earr \right\vert,
\eeq
which expresses the scalar 5-point function $E_{0}$ in terms of
five scalar 4-point functions
\beq\label{E0redf}
 E_0 = \frac{1}{\det(Y)} \, \sum_{i=0}^4 \,\det(Y_i)\,D_0(i),
\eeq
where $Y=(Y_{ij})$,
and $Y_i$ is obtained from $Y$ by replacing all
entries in the $i$th column with~1. 

In the special case of an infrared singular 5-point function we have 
\beq
Y_{00} = 2\la^2 \to 0, \qquad Y_{01} = m_1^2-p_1^2 =0, 
\qquad Y_{04} =m_4^2-p_4^2 =0,
\eeq
and the determinants fulfil the relations
\beqar\label{reldetY}
0 &=& \det(Y) - Y_{02}\det(Y_2) - Y_{03}\det(Y_3), \\
0 &=& (Y_{24}-Y_{02})\det(Y) + Y_{02}Y_{14}\det(Y_1)
        + (Y_{02}Y_{34}-Y_{03}Y_{24})\det(Y_3)
       + Y_{02}Y_{44}\det(Y_4), \nl
0 &=& (Y_{31}-Y_{03})\det(Y) + Y_{03}Y_{11}\det(Y_1)
        + (Y_{03}Y_{21}-Y_{02}Y_{31})\det(Y_2)
       + Y_{03}Y_{41}\det(Y_4), \nn
\eeqar
which allow the simplification of \refeq{E0redf}.

\subsubsection{Reduction of real 5-point functions}
 
The real 3-, 4-, and 5-point functions are defined as
\beqar \label{realCDE0}
\Cbr_0(p_1,p_2,\la,m_1,m_2) &=&
\left.\frac{2}{\pi}\int_{q_0<\De E}\!\frac{\rd^{3}{\bf q}}{2q_0}\,
\frac{1}{N'_1 N'_2}\right|_{q_0=\sqrt{{\bf q}^2+\la^2}},
\nl
\Dbr_0(p_1,p_2,p_3,\la,m_{1},m_2,m_3) &=&
\left.\frac{2}{\pi}\int\!\frac{\rd^{3}{\bf q}}{2q_0}\,
\frac{1}{N'_1 N'_2 N'_3}\right|_{q_0=\sqrt{{\bf q}^2+\la^2}},
\nl
\Ebr_0(p_1,p_2,p_3,p_4,\la,m_1,m_2,m_3,m_4) &=&
\left.\frac{2}{\pi}\int\!\frac{\rd^{3}{\bf q}}{2q_0}\,
\frac{1}{N'_1 N'_2 N'_3 N'_4}\right|_{q_0=\sqrt{{\bf q}^2+\la^2}},
\eeqar
with 
\beq \label{D0N'i}
N'_{i}= 2qp_i+p_{i}^{2}-m_{i}^{2}, \qquad i=1,\ldots,4.
\eeq
The shift of the integration boundary of $q_0$ to infinity leads
to an artificial UV divergence in the 3-point function $\Cbr_0$,
which is regularized by an energy cutoff \mbox{$\De E\to\infty$}. In
the following we only need differences of 3-point functions 
that are independent of $\De E$ and Lorentz-invariant.

Because of the appearance of UV-singular integrals in intermediate steps,
the reasoning of the previous section cannot directly be applied to $\Ebr_0$.
Therefore, we rewrite the real 5-point function as an
integral over a closed anticlockwise contour $\C$ in the $q_0$ plane and 
introduce a Lorentz-invariant UV regulator $\La$:
\beq \label{auxint}
\Ebr_0(p_1,p_2,p_3,p_4,\la,m_1,m_2,m_3,m_4)= \lim_{\La\to\infty}
\frac{1}{\ri\pi^{2}}\int_{\C}\rd^{4}q\,\frac{1}{N'_0N'_{1}\cdots N'_{4}}
\frac{-\La^2}{q^2-\La^2}
\eeq
with 
\beq
N'_0 = q^2-\la^2.
\eeq
The contour $\C$ is chosen such that it includes the poles at 
$q_0=\sqrt{\bq^2+\la^2}$ and $q_0=\sqrt{\bq^2+\La^2}$, but none else
(see \reffi{contourC})%
\footnote{\label{fn:powercount}
It is straightforward to check that the naive power counting 
for the UV behaviour in \refeq{auxint} is valid for time-like momenta $p_i$; 
light-like $p_i$ can be treated as a limiting case.
The basic idea for the proof is to deform the contour $\C$ to the vertical
line $\Re\{q_0\}=|\bq|-\epsilon$ with a small $\epsilon>0$, which is
allowed for sufficiently large $|\bq|$, more precisely when 
all particle poles appear left from the line $\Re\{q_0\}=|\bq|-\epsilon$.}.
\bfi
\centerline{
\begin{picture}(400,200)(0,0)
\LongArrow(20,100)(380,100)
\LongArrow(200,0)(200,200)
\put(375,105){$\scriptstyle \Re\{q_0\}$}
\put(210,190){$\scriptstyle \Im\{q_0\}$}
\Vertex( 50,100){2}
\Vertex(150,100){2}
\Vertex(250,100){2}
\Vertex(350,100){2}
\put( 25,80){$\scriptstyle -\sqrt{\bq^2+\La^2}$}
\put(125,80){$\scriptstyle -\sqrt{\bq^2+\la^2}$}
\put(248,80){$\scriptstyle  \sqrt{\bq^2+\la^2}$}
\put(330,80){$\scriptstyle  \sqrt{\bq^2+\La^2}$}
\CArc(250,100)(10,180,0)
\CArc(350,100)(10,180,0)
\ArrowArc(300,100)(60,0,180)
\ArrowArc(300,100)(-40,180,0)
\put(345,150){$\scriptstyle \C$}
\DashLine(245,20)(245,180){4}
\put(225,10){$\scriptstyle \Re\{q_0\}=|\bq|-\epsilon$}
\GCirc(295, 95){2}{1}
\GCirc(220,130){2}{1}
\GCirc(180,100){2}{1}
\GCirc(140, 50){2}{1}
\put(285,105){$\scriptstyle q_0(p_1)$}
\put(210,140){$\scriptstyle q_0(p_4)$}
\put(170,110){$\scriptstyle q_0(p_3)$}
\put(130, 40){$\scriptstyle q_0(p_2)$}
\end{picture} 
}
\caption{Illustration of the contour $\C$ of \refeq{auxint} in the 
complex $q_0$ plane. The open circles indicate the ``particle poles'' 
located at $q_0(p_i)=(2\bq{\bf p}_i-p_i^2+m_i^2)/(2p_{i0})$.}
\label{contourC}
\efi

The integral \refeq{auxint} can be reduced similarly to the virtual
5-point function.  
Owing to the different propagators $N'_i$, \refeq{detid} leads to
\beq\label{rE0red0}
0 = \lim_{\La\to\infty}\frac{1}{\ri\pi^2}\int_{\C}\rd^{4}q\,
\frac{1}{N'_{0}N'_{1}\cdots N'_{4}}\frac{-\La^2}{q^2-\La^2}
\left\vert
\barr{cccc}
2N'_{0}  & 2qp_{1}    & \ldots & \;2qp_{4} \\
N'_{1}+Y_{10} & 2p_{1}^{2} & \ldots & \;2p_{1}p_{4} \\
\vdots    & \vdots     & \ddots     &\;\vdots     \\
N'_{4}+Y_{40} & 2p_{4}p_{1} & \ldots &\; 2p_{4}^{2}
\earr
\right\vert
\eeq
instead of \refeq{E0red0}, with $Y_{ij}$ from \refeq{defY},
and $\lambda^2$ can be set to zero in all $Y_{ij}$, in particular we
have $Y_{00} = 2\la^2 \to 0$. After expanding the determinant along the 
first column and using the Lorentz
decompositions of the integrals where $N'_{k}$ is cancelled, we see
that these terms vanish, and we are left with
\beqar\label{rE0red2}
0&=&\lim_{\La\to\infty}\frac{1}{\ri\pi^2}\int_{\C}\rd^{4}q\,
\frac{1}{N'_{0}N'_{1}\cdots N'_{4}}\frac{-\La^2}{q^2-\La^2}
\left\vert
\barr{cccc}
2N'_{0}  & 2qp_{1}    & \ldots & \;2qp_{4} \\
Y_{10} & 2p_{1}^{2} & \ldots & \;2p_{1}p_{4} \\
\vdots    & \vdots     & \ddots     &\;\vdots     \\
Y_{40} & 2p_{4}p_{1} & \ldots &\; 2p_{4}^{2}
\earr
\right\vert.
\eeqar
Using
\beqar
2p_{i}p_{j}&=& Y_{ij}-Y_{i0}-Y_{0j},\nl
2qp_{j}&=& N'_{j}+Y_{0j},
\eeqar
adding the first column to the other columns and extending the
determinant leads to
\beq
0=\lim_{\La\to\infty}\frac{1}{\ri\pi^2}\int_{\C}\rd^{4}q\,
\frac{1}{N'_{0}N'_{1}\cdots N'_{4}}\frac{-\La^2}{q^2-\La^2}
\left\vert
\barr{ccccc}
1  & 0  &  Y_{01}   & \ldots & \; Y_{04} \\      
0 & 2N'_{0} & N'_{1}+2N'_0+Y_{01}    & \ldots & \;N'_4+2N'_0+Y_{04} \\
0 & Y_{10} & Y_{11}-Y_{01} & \ldots & \; Y_{14}-Y_{04} \\
\vdots & \vdots    & \vdots     & \ddots     &\;\vdots     \\
0 & Y_{40} &  Y_{41}-Y_{01}& \ldots &\;  Y_{44}-Y_{04}
\earr
\right\vert.\nl
\eeq
Subtracting the first row from the second, adding the first row to the
other rows, and exchanging the first two rows, we arrive at
\beq\label{almostE0brred}
0=\lim_{\La\to\infty}\frac{1}{\ri\pi^2}\int_{\C}\rd^{4}q\,
\frac{1}{N'_{0}N'_{1}\cdots N'_{4}}\frac{-\La^2}{q^2-\La^2}
\left\vert
\barr{ccccc}
-1 & 2N'_{0} & N'_{1}+2N'_0    & \ldots & \;N'_4+2N'_0 \\
1  & 0  &  Y_{01}   & \ldots & \; Y_{04} \\      
1 & Y_{10} & Y_{11} & \ldots & \; Y_{14} \\
\vdots & \vdots    & \vdots     & \ddots     &\;\vdots     \\
1 & Y_{40} &  Y_{41}& \ldots &\;  Y_{44}
\earr
\right\vert.
\eeq
Now we perform the contour integral
using power counting for $\Lambda\to\infty$ (see
footnote~\ref{fn:powercount}).
In the contribution of the pole at
$q_0=\sqrt{\bq^2+\la^2}$ we have $N'_0=0$ in the numerator;
the limit $\Lambda\to\infty$ can be trivially taken and reproduces usual
bremsstrahlung integrals, as defined in \refeq{realCDE0}.
In the contribution of the pole at $q_0=\sqrt{\bq^2+\La^2}$ the term containing 
$N'_0$ in the numerator survives and will be calculated below, 
but all other terms vanish after taking the limit $\La\to\infty$. 

Thus, we find
\beq\label{E0brred}
0 \;=\;
\left|\begin{array}{cccccc}
\;-\Ebr_0\; &\; \tilde\Dbr_0(0)\; 
&\; \Dbr_0(1) +\tilde\Dbr_0(0)\; &\; \Dbr_0(2) +\tilde\Dbr_0(0)\;
&\; \Dbr_0(3) +\tilde\Dbr_0(0)\; &\; \Dbr_0(4) +\tilde\Dbr_0(0)\; \\
1 & Y_{00} & Y_{01} & Y_{02} & Y_{03} & Y_{04} \\
1 & Y_{10} & Y_{11} & Y_{12} & Y_{13} & Y_{14} \\
1 & Y_{20} & Y_{21} & Y_{22} & Y_{23} & Y_{24} \\
1 & Y_{30} & Y_{31} & Y_{32} & Y_{33} & Y_{34} \\
1 & Y_{40} & Y_{41} & Y_{42} & Y_{43} & Y_{44}
\end{array}\right|,
\eeq
or
\beq\label{Erred}
\Ebr_0 = \frac{1}{\det(Y)} \,
\biggl\{\det(Y_0)\,\tilde\Dbr_0(0)
+\sum_{i=1}^4 \,\det(Y_i)\,\left[\Dbr_0(i)+\tilde\Dbr_0(0)\right]\biggr\}.
\eeq
Here 
\beq
\tilde\Dbr_0(0) = \lim_{\La\to\infty}\frac{2}{\ri\pi^2}\int_\C\rd^{4}q\,
\frac{1}{N'_{1}\cdots N'_{4}}\frac{-\La^2}{q^2-\La^2},
\eeq
and the 4-point bremsstrahlung integrals $\Dbr_0(i)$, $i = 1,2,3,4$,
result from $\Ebr_0$ by omitting the $i$th denominator $N'_i$.
The result \refeq{E0brred} differs from \refeq{E0red2} only by the 
extra $\tilde\Dbr_0(0)$'s added to the $\Dbr_0(i)$'s.

The integral $\tilde\Dbr_0(0)$ stems from the terms involving $N'_0$
in the numerator in \refeq{almostE0brred} and can be expressed as follows:
\beqar
\tilde\Dbr_0(0) 
 &=&\frac{2}{\ri\pi^2}\int_{\C'}\rd^{4}q'\,
\frac{1}{2q'p_1\cdots 2q'p_4}\frac{-1}{q'^2-1}, 
\eeqar
where the contour $\C'$ surrounds $q'_0=\sqrt{{\bf q'}^2+1}$.
Performing the contour integral over $\rd q'_0$ yields
\beqar
\tilde\Dbr_0(0) &=& \frac{4}{\pi}
\int\frac{\rd^{3}{\bf q'}}{2q'_0}\,
\frac{1}{2q' p_1\cdots 2q' p_4}
\biggr|_{q'_0=\sqrt{{\bf q'}^2+1}}.
\eeqar
Now the vector $q'$ is time-like. Since also the vectors $p_i$ are 
time-like (or at least light-like), the scalar products 
$q'p_i$ cannot become zero. After redefining the momenta,
\beq\label{eq:momredef}
p_i = \si_i \tilde p_i,  \qquad \si_i = \pm 1, \qquad \tilde p_{i0}> 0,
\eeq
and extracting the signs $\si_i$, 
this integral can be evaluated by a Feynman-parameter representation and 
momentum integration in polar coordinates resulting in:
\beqar
\tilde\Dbr_0(0) &=& 
 - \si_1\si_2\si_3\si_4\int_0^\infty\rd x_1\,\rd x_2\,\rd x_3\,\rd x_4\, 
\de\biggl(1-\sum_{i=1}^4 x_i\biggr)
\left[\biggl(\sum_{i=1}^4 x_i \tilde p_i\biggr)^2\right]^{-2}.
\eeqar
This is just the Feynman-parameter representation of a virtual 4-point
function such that we finally obtain
\beq
\tilde\Dbr_0(0) =- \si_1\si_2\si_3\si_4 D_0\left(\tilde p_2-\tilde p_1,\tilde
  p_3-\tilde p_1,\tilde p_4-\tilde p_1,
\sqrt{p_1^2},\sqrt{p_2^2},\sqrt{p_3^2},\sqrt{p_4^2}\right).
\eeq 

\subsubsection{\boldmath{Explicit reduction of the virtual 5-point function 
for the photon exchange between $\bar f_2$ and $f_3$}}

For the photon exchange between $\bar f_2$ and $f_3$ 
the following scalar integrals are relevant:
\beqar\label{virt_ints}
E_0 &=& E_0(-k_3,-k_-,k_+,k_2,\la,m_3,M,M,m_2),
\nl
D_0(0) &=& D_0(-k_4,k_+ +k_3,k_2+k_3,0,M,M,0),
\nl
D_0(1) &=& D_0(-k_-,k_+,k_2,0,M,M,m_2),
\nl
D_0(2) &=& D_0(-k_3,k_+,k_2,\la,m_3,M,m_2),
\nl
D_0(3) &=& D_0(-k_3,-k_-,k_2,\la,m_3,M,m_2)
\nlc
D_0(4) &=& D_0(-k_3,-k_-,k_+,0,m_3,M,M).
\eeqar
Since we neglect the external fermion masses, the last two relations
\refeq{reldetY} simplify to
\beqar\label{reldetYipole}
0 &=& (s_{23}+s_{24})\det(Y) + K_-s_{23}\det(Y_1)
        - \left[ K_+(s_{23}+s_{24})+K_-M_+^2 \right]\det(Y_3), \nl
0 &=& (s_{13}+s_{23})\det(Y) + K_+s_{23}\det(Y_4)
        - \left[ K_-(s_{13}+s_{23})+K_+M_-^2 \right]\det(Y_2).
\eeqar
These relations allow 
us to eliminate $\det(Y_1)$ and $\det(Y_4)$ from \refeq{E0redf}, resulting in:
\begin{eqnarray}\label{vE5redpole}
\nonumber
\lefteqn{E_0(-k_3,-k_-,k_+,k_2,\lambda,m_3,M,M,m_2)=
\frac{\det(Y_0)}
{\det(Y)} D_0(0)}\quad\\
\nonumber
&&{}+ \frac{\det(Y_3)}{\det(Y)K_-s_{23}}
\Big\{[K_+(s_{23}+s_{24})+K_-M_+^2]D_0(1)+K_-s_{23}D_0(3)\Big\}\\
\nonumber
&&{}+ \frac{\det(Y_2)}{\det(Y)K_+s_{23}}
\Big\{[K_-(s_{13}+s_{23})+K_+M_-^2]D_0(4)+K_+s_{23}D_0(2)\Big\}\\
&&
-\frac{s_{13}+s_{23}}{K_+s_{23}} D_0(4)
-\frac{s_{23}+s_{24}}{K_-s_{23}} D_0(1).
\end{eqnarray}
The matrix $Y$ reads
\beqar\label{Yvpole}
Y &=& \left(\begin{array}{@{\ }c@{\ \ }c@{\ \ }c@{\ \ }c@{\ \ }c@{\ }}
{}0      & 0       & -K_-     & -K_+       & 0 \\
{}*      & 0       &  M^2     & \;(-K_+-s_{13}-s_{23})\; & -s_{23} \\
{}*      & *       & 2M^2     & \; 2M^2-s\;& (-K_- -s_{23}-s_{24}) \\
{}*      & *       & *        & 2M^2       & M^2 \\
{}*      & *       & *        & *          & 0 
\end{array}\right).
\eeqar
Neglecting terms that do not contribute to the correction factor in
DPA, the corresponding determinants are given by
\beqar\label{detYvpole}
\det(Y) &\sim& 2s_{23} \left[
K_+K_-s_{14}s_{23} - (K_+\MW^2+K_-s_{13})(K_-\MW^2+K_+s_{24}) \right],
\nl
\det(Y_0) &\sim& -\kappa_\PW^2,
\nl
\det(Y_1)
&\sim& K_+\left[\MW^4(s_{23}-s_{24})
                +(s_{23}+s_{24})(s_{13}s_{24}-s_{14}s_{23})\right]
\nn\\ && {}
+ K_-\MW^2(-\MW^4+2s_{13}s_{23}+s_{13}s_{24}+s_{14}s_{23}),
\nl
\det(Y_2)
&\sim& -s_{23}\left[ K_+(\MW^4+s_{13}s_{24}-s_{14}s_{23})
                +2K_-\MW^2s_{13} \right],
\nl
\det(Y_3) &=& \det(Y_2)\big|_{K_+\leftrightarrow K_-,
                s_{13}\leftrightarrow s_{24}},
\nl
\det(Y_4) &=& \det(Y_1)\big|_{K_+\leftrightarrow K_-,
                s_{13}\leftrightarrow s_{24}},
\eeqar
where the 
shorthand $\kappa_\PW$ is defined in \refapp{prelim}.

\subsubsection{\boldmath{Explicit reduction of the real 5-point function for 
the photon exchange between $\bar f_2$ and $f_3$}}

In this case the integrals appearing in \refeq{Erred} read
\beqar\label{real_ints}
\Ebr_0 &=& \Ebr_0(k_3,k_-,k_+,k_2,\la,m_3,M^*,M,m_2),
\nl
\tilde\Dbr_0(0) &=& 
-D_0(k_4,k_+-k_3,k_2-k_3,0,M_-,M_+,0),
\nl
\Dbr_0(1) &=& \Dbr_0(k_-,k_+,k_2,0,M^*,M,m_2),
\nl
\Dbr_0(2) &=& \Dbr_0(k_3,k_+,k_2,\la,m_3,M,m_2),
\nl
\Dbr_0(3) &=& \Dbr_0(k_3,k_-,k_2,\la,m_3,M^*,m_2)
\nlc
\Dbr_0(4) &=& \Dbr_0(k_3,k_-,k_+,0,m_3,M^*,M).
\eeqar
In analogy to the virtual  5-point function, we can express the real
5-point function with the help of the relations (we denote the matrix $Y$ for 
the real 5-point function with a prime, in order to distinguish it from the 
one for the virtual 5-point function):
\beqar
0 &=&
(s_{23}+s_{24})\det(\Ybr) + K_-^*s_{23}\det(\Ybr_1)
        - \left[ K_+(s_{23}+s_{24})-K_-^*M_+^2 \right]\det(\Ybr_3), \nl
0 &=&
(s_{13}+s_{23})\det(\Ybr) + K_+s_{23}\det(\Ybr_4)
        - \left[ K_-^*(s_{13}+s_{23})-K_+M_-^2 \right]\det(\Ybr_2),
\eeqar
by
\begin{eqnarray}\label{rE5redpole}
\lefteqn{\Ebr_0(k_3,k_-,k_+,k_2,\la,m_3,M^*,M,m_2) =
\frac{\det(\Ybr_0)}{\det(\Ybr)} \tilde\Dbr_0(0)}\nl
\nonumber
&&{}+ \frac{\det(\Ybr_3)}{\det(\Ybr)K_-^*s_{23}}
\biggl\{\Bigl[K_+(s_{23}+s_{24})-K_-^*M_+^2\Bigr]
\Bigl[\Dbr_0(1)+\tilde\Dbr_0(0)\Bigr]
+K_-^*s_{23}\Bigl[\Dbr_0(3)+\tilde\Dbr_0(0)\Bigr]\biggr\}\\
\nonumber
&&
{}+ \frac{\det(\Ybr_2)}{\det(\Ybr)K_+s_{23}}
\biggl\{\Bigl[K_-^*(s_{13}+s_{23})-K_+M_-^2\Bigr]
\Bigl[\Dbr_0(4)+\tilde\Dbr_0(0)\Bigr]
+K_+s_{23}\Bigl[\Dbr_0(2)+\tilde\Dbr_0(0)\Bigr]\biggr\}\\
&&{}
-\frac{s_{13}+s_{23}}{K_+s_{23}} \Bigl[\Dbr_0(4)+\tilde\Dbr_0(0)\Bigr]
-\frac{s_{23}+s_{24}}{K_-^*s_{23}} \Bigl[\Dbr_0(1)+\tilde\Dbr_0(0)\Bigr].
\end{eqnarray}
For the matrix $\Ybr$ we find
\beqar
\Ybr &=& \left(\begin{array}{ccccc}
\ 0 & 0 & -K_-^* & -K_+                   & 0 \\
\ * & 0 & (M^*)^2  & (-K_++s_{13}+s_{23}) & s_{23} \\
\ * & * & 2(M^*)^2 & \;(-2K_+-2K_-^*+s-(M^*)^2-M^2)\; 
                                          & (-K_-^*+s_{23}+s_{24}) \\
\ * & * &   *    & 2M^2                   & M^2 \\
\ * \ & \ * \ & \ * \ & \ * \ & 0
\end{array}\right).
\nln
\eeqar
Replacing $-K_-^*$  by $K_-$ and $(M^*)^2$ by $M^2$ and multiplying the 
second and third columns and rows by $-1$, this becomes equal to  
\refeq{Yvpole} in DPA.

In DPA, $\tilde\Dbr_0(0)$ can be neglected in the terms
$\Dbr_0(i)+\tilde\Dbr_0(0)$ in \refeq{Erred} and \refeq{rE5redpole},
and the reduction of the real 5-point function becomes 
algebraically identical to the reduction of the virtual 5-point function,
apart from the differences in signs of some momenta.
As a consequence, the results for the virtual corrections can be translated
to the real case if we substitute $K_-\to -K_-^*$ in all 
algebraic factors such as the determinants and $E_0\to\Ebr_0$,
$D_0(0)\to \tilde\Dbr_0(0)$, 
$D_0(1)\to -\Dbr_0(1)$, $D_0(2)\to -\Dbr_0(2)$, $D_0(3)\to
\Dbr_0(3)$ and $D_0(4)\to\Dbr_0(4)$.
In particular, the determinants are related by
\beqar\label{eq:detrels}
\det(\Ybr) &\sim& +\det(Y)\Big|_{K_-\to -K_-^*} ,\qquad
\det(\Ybr_0) \sim +\det(Y_0) \sim -\kappa_\PW^2                \nlc
\det(\Ybr_1) &\sim& -\det(Y_1)\Big|_{K_-\to -K_-^*},\qquad 
\det(\Ybr_2) \sim -\det(Y_2)\Big|_{K_-\to -K_-^*} \nlc
\det(\Ybr_3) &\sim& +\det(Y_3)\Big|_{K_-\to -K_-^*}, \qquad
\det(\Ybr_4) \sim +\det(Y_4)\Big|_{K_-\to -K_-^*}. 
\eeqar

\subsection{Calculation of 4- and 3-point functions}

The scalar loop integrals have been evaluated following the methods 
of \citere{tH79}. Our explicit results are listed in \refapp{appvirtual}.
For vanishing \PW-boson width they agree with the general results
of \citere{tH79}. For finite \PW-boson width the virtual 4-point functions 
are in agreement with those of \citeres{Be97a,Be97b}
in DPA. This shows explicitly that the $q^2$ terms in the
W-boson and fermion propagators are irrelevant in DPA. 

An evaluation of the bremsstrahlung integrals, which follows
closely the techniques for calculating loop integrals, is sketched in
\refapp{appbrcal}. 
The final results in DPA are listed in \refapp{appreal}, and the 4-point
functions agree with those of \citeres{Be97a,Be97b}.
We have analytically reproduced all exact results for the 
occurring bremsstrahlung 4- and 3-point integrals 
by independent methods.
In addition, we have evaluated the IR-finite integrals $\Dbr_0(1)$,
$\Dbr_0(4)$, and $\Ebr_0-\Dbr_0(3)/K_+$ by a direct 
Monte Carlo integration over the photon momentum,
yielding perfect agreement with our exact
analytical results for these integrals. Note that this, in particular,
checks the reduction of the bremsstrahlung 5-point function
described in the previous section.

\begin{sloppypar}
Our results for the 3-point functions cannot directly be compared
with those of \citeres{Be97a,Be97b}, 
because different approaches have been used.
While our results are IR-singular owing to the subtracted on-shell
integrals, the results of \citeres{Be97a,Be97b} are 
artificially UV-singular owing to the neglect of $q^2$ in the 
W~propagators. However, when adding the real and virtual 3-point functions 
the two results agree. This confirms that
our definition of the non-factorizable corrections is equivalent to the
one of \citeres{Be97a,Be97b}
in DPA. Thus, it turns out that in DPA the subtraction of the on-shell
contribution is effectively equivalent to the neglect of the $q^2$ terms
in all but the photon propagators.
\end{sloppypar}

\section{Analytic results for the non-factorizable corrections}
\label{seanres}

\subsection{General properties  of non-factorizable corrections}

In \citere{Me96} it was shown from the integral representation that the
non-factorizable corrections associated with 
photon exchange between initial and final state vanish in DPA. This was
confirmed in \citeres{Be97a,Be97b}. Via explicit evaluation of all integrals
we have checked that the cancellation between virtual and real
integrals takes place 
diagram by diagram once the factorizable contributions are subtracted.
In this way all interference terms (if), (mf), (im), and (mm) drop
out. Examples for the virtual Feynman diagrams contributing to these
types of corrections are shown in \reffi{virtual_initial_final_diagrams}.

The only non-vanishing non-factorizable corrections are due to the 
contributions (\ffp), (\mfp), and (\mmp). The corresponding virtual
diagrams are shown in \reffi{virtual_final_final_diagrams},
apart from permutations of the final-state fermions. Two of the 
corresponding real diagrams are pictured in
\reffis{real_non_factorizable_diagram} and \ref{real_mixed_diagram}.
Since these corrections depend only on $s$-channel invariants, 
the non-factorizable corrections are independent of the
production  angle of the \PW~bosons,
as was also pointed out in \citeres{Be97a,Be97b}.

\subsection{Generic form of the correction factor}
\label{gfcf}

The non-factorizable corrections $\rd\si_\nf$ to the fully differential 
lowest-order cross-section $\rd\si_\born$ resulting from the matrix element
\refeq{mborn} take the form of a correction factor to the
lowest-order cross-section:
\beq
\rd \si_\nf = \de_\nf \, \rd\si_\born.
\eeq
Upon splitting the contributions that result from photons coupled to
the \PW~bosons according to
$1=Q_\PWp=Q_1-Q_2$ and $1=-Q_\PWm=Q_4-Q_3$ into contributions
associated with definite final-state fermions, the
complete correction factor to the lowest-order cross-section can be written as
\beq\label{nfcorrfac}
\delta_\nf = \sum_{a=1,2} \, \sum_{b=3,4} \, (-1)^{a+b+1} \, Q_a Q_b \,
\frac{\alpha}{\pi} \, \Re\{\Delta(k_+,k_a;k_-,k_b)\}.
\eeq
In the following, only $\Delta=\Delta(k_+,k_2;k_-,k_3)$ is given; 
the other terms follow by obvious substitutions. As discussed above, $\De$
gets contributions from {\it intermediate--intermediate} 
($\Delta_{\mmp}$), {\it intermediate--final}
($\Delta_{\mfp}$), and {\it final--final}
($\Delta_{\ffp}$) interactions:
\beq\label{eq:Delta}
\Delta = \Delta_{\mmp}+\Delta_{\mfp}+\Delta_{\ffp}.
\eeq
The quantity ($\Delta_{\mmp}$ is independent of the final-state
fermions. The individual contributions read
\beqar
\label{eq:dmm}
\Delta_{\mmp} &\sim& (2\MW^2-s)\biggl\{
C_0(k_+,-k_-,0,M,M)
- \Bigl[C_0(k_+,-k_-,\la,\MW,\MW)\Bigr]_{k_\pm^2=\MW^2} 
\nn\\ && {}
\phantom{(2\MW^2-s)\biggl[}
- \Cbr_0(k_+,k_-,0,M,M^*) 
+ \Bigl[\Cbr_0(k_+,k_-,\la,\MW,\MW)\Bigr]_{k_\pm^2=\MW^2} \biggr\},
\hspace{2em}
\\[.5em]
\Delta_{\mfp} &\sim& 
- (s_{23}+s_{24})\Bigl[ K_+D_0(1)-K_+\Dbr_0(1) \Bigr]
- (s_{13}+s_{23})\Bigl[ K_-D_0(4)-K_-^*\Dbr_0(4) \Bigr],
\\[.5em]
\Delta_{\ffp} &\sim& 
-K_+s_{23}\Bigl[ K_-E_0 -K_-^*\Ebr_0\Bigr],
\eeqar
with the arguments of the 5- and 4-point functions as defined in
\refeq{virt_ints} and \refeq{real_ints}. 

The sum $\Delta_{\mfp}+\Delta_{\ffp}$ can be
simplified by inserting the decompositions of the 5-point functions
\refeq{vE5redpole} and \refeq{rE5redpole}. In DPA this leads to
\beqar\label{denf1}
\Delta_{\mfp}+\Delta_{\ffp} &\sim&
-\frac{K_+K_-s_{23}\det(Y_0)}{\det(Y)}D_0(0)
+\frac{K_+K_-^*s_{23}\det(\Ybr_0)}{\det(\Ybr)}\tilde\Dbr_0(0)
\nn\\ && {}
-\frac{K_+\det(Y_3)}{\det(Y)}
\left\{[K_+(s_{23}+s_{24})+K_-\MW^2]D_0(1) + K_-s_{23}D_0(3)\right\} 
\nn\\ && {}
-\frac{K_-\det(Y_2)}{\det(Y)}
\left\{[K_-(s_{13}+s_{23})+K_+\MW^2]D_0(4) + K_+s_{23}D_0(2)\right\}
\nn\\ && {}
+ \frac{K_+\det(\Ybr_3)}{\det(\Ybr)}
\left\{ [K_+(s_{23}+s_{24})-K_-^*\MW^2]\Dbr_0(1) + K_-^*s_{23}\Dbr_0(3) 
\right\}
\nn\\ && {}
+ \frac{K_-^*\det(\Ybr_2)}{\det(\Ybr)}
\left\{ [K_-^*(s_{13}+s_{23})-K_+\MW^2]\Dbr_0(4) + K_+s_{23}\Dbr_0(2) 
\right\}.
\eeqar
Note that $\Delta_{\mfp}$ is exactly cancelled by the contributions 
of the last two terms in \refeq{vE5redpole} and \refeq{rE5redpole}. 

Inserting the expressions for the scalar integrals from \refapp{appscalint}
and using the first of the relations \refeq{reldetY},
various terms, notably all IR divergences and mass-singular logarithms, 
cancel between the real and virtual corrections, and in DPA we are left with
\begin{eqnarray}
\label{eq:dmmDPA}
\Delta_{\mmp} &\sim& 
\frac{2\MW^2-s}{s\beta_\PW}\biggl[
-\cLi\biggl(\frac{K_-}{K_+},x_\PW\biggr)
+\cLi\biggl(\frac{K_-}{K_+},x_\PW^{-1}\biggr)
+\cLi\biggl(-\frac{K_-^*}{K_+},x_\PW\biggr)
\nn\\ && {}
-\cLi\biggl(-\frac{K_-^*}{K_+},x_\PW^{-1}\biggr)
+ 2\pi\ri \ln\biggl(\frac{K_+ +K_-^*x_\PW}{\ri\MW^2}\biggr)
\biggr]
\; + \; \mbox{imaginary parts,}
\hspace{2em}
\label{eq:DmmDPA}
\\[1em]
\Delta_{\mfp}+\Delta_{\ffp} &\sim &
-\frac{K_+K_-s_{23}\det(Y_0)}{\det(Y)}D_0(0)
- \frac{K_+ \det(Y_3)}{\det( Y)} F_3
- \frac{K_- \det(Y_2)}{\det( Y)} F_2
\nn\\ && {}
+\frac{K_+K_-^*s_{23}\det(\Ybr_0)}{\det(\Ybr)}\tilde\Dbr_0(0)
+ \frac{\Kp \det(\Ybr_3)}{\det( \Ybr)} \F_3
+ \frac{\Km \det(\Ybr_2)}{\det( \Ybr)} \F_2
\nn\\ && {}
+ \; \mbox{imaginary parts,}
\label{eq:DmfDPA}
\end{eqnarray}
with $D_0(0)$ and $\tilde\Dbr_0(0)$ 
given in \refeq{D00} and \refeq{Dbr00}, respectively, and
\beqar\label{Fi}
F_3&=&
-2\cLi\biggl(\frac{K_+}{K_-},-\frac{s_{23}+s_{24}}{M_W^2}-\ri\epsilon\biggr)
+\sum\limits_{\tau=\pm 1}\biggl[
\cLi\biggl(\frac{K_+}{K_-},\xW^\tau\biggr)-
\cLi\biggl(-\frac{M_W^2}{s_{23}+s_{24}}+\ri\epsilon,\xW^\tau\biggr)\biggr]
\nn\\ && {}
-\Li\biggl(-\frac{s_{24}}{s_{23}}\biggr)
+2\ln\biggl(-\frac{s_{23}}{M_W^2}-\ri \epsilon\biggr)
  \ln\biggl(-\frac{K_-}{M_W^2}\biggr)
- \ln^2\biggl(-\frac{s_{23}+s_{24}}{M_W^2}-\ri \epsilon\biggr),
\nl[1em]
\F_3&=&
F_3\Big|_{K_- \to -K_-^*}
+ 2\pi\ri \biggl[
2\ln\biggl(1-\frac{K_+}{K_-^*}\frac{s_{23}+s_{24}}{\MW^2}\biggr)
-\ln\biggl(1+\frac{K_+}{K_-^* x_\PW}\biggr)
-\ln\biggl(1+\frac{x_\PW \MW^2}{s_{23}+s_{24}}\biggr)
\nn\\ && {}
\qquad\qquad\qquad
+\ln\biggl(\frac{\ri K_-^*}{s_{23}+s_{24}}\biggr)
\biggr],
\nl[1em]
F_2&=& F_3\Big|_{K_+\leftrightarrow K_-,s_{24}\leftrightarrow s_{13}}, 
\nl[1em]
\F_2&=& F_2\Big|_{K_- \to -K_-^*}
+ 2\pi\ri \biggl\{
\ln\biggl(1+\frac{K_-^*x_\PW}{K_+}\biggr) 
-\ln\biggl(1+\frac{x_\PW \MW^2}{s_{13}+s_{23}}\biggr)
+\ln\biggl[\frac{K_+s_{23}}{\ri\MW^2(s_{13}+s_{23})}\biggr]
\biggr\}. 
\nn\\
\end{eqnarray}
The variables $\beta_\PW$, $x_\PW$ and the dilogarithms $\Li$,
$\cLi$ are defined in \refapp{prelim}.

The above results contain logarithms and dilogarithms
the arguments of which depend on $K_\pm$. It is interesting to see whether 
enhanced logarithms of the form $\ln (K_\pm/\MW^2)$ appear
in the limits $K_\pm\to0$. It turns out that such logarithms are absent 
from the non-factorizable corrections,
irrespective of the ratio in which the two limits $K_\pm\to0$ are realized.

\subsection{Inclusion of the exact off-shell Coulomb singularity}
\label{se43}

For non-relativistic \PW~bosons, \ie for 
a small on-shell velocity $\be_\PW$, the long range of
the Coulomb interaction leads to a large radiative correction, known as
the Coulomb singularity. For on-shell \PW~bosons,
this correction behaves like $1/\be_\PW$
near threshold, but including the instability of the 
\PW~bosons the long-range interaction is effectively truncated,
and the $1/\beta_\PW$ singularity is regularized. Therefore, for realistic
predictions in the threshold region, the on-shell Coulomb singularity
should be replaced by the corresponding off-shell correction.
The precise form of this off-shell Coulomb singularity \cite{Coul} reveals that 
corrections of some per cent occur even a few \PW-decay widths above threshold.
As explained above, the DPA becomes valid only 
several widths above threshold. Nevertheless, there exists an overlapping 
region in which the inclusion of the Coulomb singularity within the 
double-pole approach is reasonable.

The Feynman graph relevant for the Coulomb singularity is diagram (d) 
of \reffi{virtual_final_final_diagrams}. The
non-factorizable corrections contain just the difference between 
the off-shell and the on-shell contributions of this diagram. Therefore,
the difference between off-shell and on-shell Coulomb singularity 
is in principle included in $\Delta_{\mathrm{mm}'}$, as defined in
\refeq{eq:dmm}. The genuine form of
$\Delta_{\mathrm{mm}'}$ in DPA, which is given by \refeq{eq:DmmDPA},
does not contain 
the full effect of the Coulomb singularity, because 
in both $C_0$ functions of \refeq{eq:dmm} the on-shell limit
$K_\pm\to 0$ was taken under the assumption of a finite $\be_\PW$.
In order to include the correct difference 
between off-shell and on-shell Coulomb singularity
in $\Delta_{\mathrm{mm}'}$, the on-shell limit of the $C_0$ functions of 
\refeq{eq:dmm} has to be taken for arbitrary $\be_\PW$. 
The full off-shell Coulomb singularity can be included by adding 
\beq\label{demmcoul}
\frac{2\MW^2-s}{s} \biggl[
\frac{2\pi\ri}{\betap} \ln\biggl(
\frac{\betaM+\Delta_M-\betap}{\betaM+\Delta_M+\betap}\biggr) 
-\frac{2\pi\ri}{\betaW} \ln\biggl(
\frac{K_+ + K_- + s\betaW\Delta_M}{2\betaW^2s}\biggr) \biggr]
\eeq
to the genuine DPA result \refeq{eq:dmmDPA} for $\Delta_{\mathrm{mm}'}$.
The quantities $\betaM$, $\betap$, and $\Delta_M$ are defined in
\refeq{abbrev}.  
After combination with the factorizable doubly-resonant corrections,
all doubly-resonant corrections and the correct off-shell Coulomb singularity
are included. The on-shell Coulomb
singularity contained in the factorizable corrections is compensated
by a corresponding contribution in the non-factorizable ones. 
Note, however, that this subtracted on-shell Coulomb singularity appears 
as an artefact if the non-factorizable corrections are discussed separately.
 
\subsection{Non-local cancellations}

In \citere{Fa94} it was pointed out that the non-factorizable
photonic corrections completely cancel in DPA if the phase-space 
integration over both invariant masses of the W~bosons is performed.
This cancellation is due to a symmetry of the non-factorizable corrections.

The lowest-order cross section in DPA
is symmetric with respect to the ``reflections'' 
\beqar
\label{reflect}
K_+=(k_+^2-\MW^2)+\ri\MW\Gamma_\PW 
&\;\leftrightarrow\;& -(k_+^2-\MW^2)+\ri\MW\Gamma_\PW=-K_+^*, \nn\\
K_-=(k_-^2-\MW^2)+\ri\MW\Gamma_\PW 
&\;\leftrightarrow\;& -(k_-^2-\MW^2)+\ri\MW\Gamma_\PW=-K_-^*. 
\eeqar
Therefore, $\De$ can be 
symmetrized in $K_+ \to -K_+^*$ or $K_- \to -K_-^*$ if the respective 
invariant mass is integrated out. For instance, if $k_-^2$ is integrated
out, $\De$ can be replaced by
\beqar
\label{kmaverage}
\lefteqn{
\frac{1}{2}\left(\Delta+\Delta\Big|_{K_- \leftrightarrow -K_-^*}\right) } && 
\nn\\*
& \;\sim\; &
\ri\pi \biggl\{ \;
\frac{s-2\MW^2}{\beta_\PW s}
\ln\biggl(\frac{K_- x_\PW+K_+^*}{K_- x_\PW-K_+}\biggr)
\nn\\ && {} \qquad
+ K_+s_{23}\kappa_\PW 
\biggl[ \frac{K_-}{\det(Y)}-\frac{K_-^*}{\det(\Ybr)} \biggr]
\biggl[ \ln\biggl(-x_\PW\frac{s_{23}}{\MW^2}\biggr)
+\ln\biggl(1+\frac{s_{13}}{s_{23}}(1-z)\biggr)
\nn\\ && {} \qquad\qquad
+\ln\biggl(1+\frac{s_{24}}{s_{23}}(1-z)\biggr) 
-\ln\biggl(1+\frac{s_{13}}{\MW^2}z x_\PW\biggr)
-\ln\biggl(1+\frac{s_{24}}{\MW^2}z x_\PW\biggr) \biggr]
\nn\\ && {} \qquad
-\biggl[\frac{K_- \det(Y_2)}{\det(Y)}
+\frac{K_-^* \det(\Ybr_2)}{\det(\Ybr)}\biggr] 
\biggl[ \ln\biggl(x_\PW+\frac{s_{13}+s_{23}}{\MW^2}\biggr)
-\ln\biggl(-x_\PW\frac{s_{23}}{\MW^2}\biggr) \biggr]
\nn\\ && {} \qquad
-\frac{K_+\det(Y_3)}{\det(Y)}\biggl[
\ln\biggl(x_\PW+\frac{s_{23}+s_{24}}{\MW^2}\biggr)
-2\ln\biggl(1+\frac{K_+}{K_-}\frac{s_{23}+s_{24}}{\MW^2}\biggr) \biggr]
\nn\\ && {} \qquad
-\frac{K_+ \det(\Ybr_3)}{\det(\Ybr)}\biggl[
\ln\biggl(x_\PW+\frac{s_{23}+s_{24}}{\MW^2}\biggr)
-2\ln\biggl(1-\frac{K_+}{K_-^*}\frac{s_{23}+s_{24}}{\MW^2}\biggr) \biggr]
\nn\\ && {} \qquad
+\frac{2s_{23}\MW^2(K_+^2 s_{24}-K_-^2 s_{13})}{\det(Y)}
\ln\biggl(1-\frac{K_+}{K_- x_\PW}\biggr)
\nn\\ && {} \qquad
+\frac{2s_{23}\MW^2(K_+^2 s_{24}-K_-^{*2} s_{13})}{\det(\Ybr)}
\ln\biggl(1+\frac{K_+}{K_-^* x_\PW}\biggr)
\; \biggr\}
\nn\\ && {} 
+ \; \mbox{imaginary parts},
\eeqar
with $z$ defined in \refeq{eq:xtilde}.
Note that this expression is considerably simpler than the full result 
for $\De$. In particular, all dilogarithms have dropped out. 

Symmetrizing \refeq{kmaverage} also in $K_+ \leftrightarrow -K_+^*$
leads to further simplifications if the relations \refeq{eq:detrels} for
the determinants are used. Under the assumptions that
$(s_{13}+s_{23})>-\MW^2 x_\PW$, $(s_{23}+s_{24})>-\MW^2 x_\PW$, 
and that $\kappa_\PW$ is imaginary, the real part of the result vanishes.
These assumptions are fulfilled on resonance, $k_\pm^2=\MW^2$; 
off resonance, there are regions in phase space 
where the assumptions are
violated. The volume of those regions of phase space is suppressed 
by factors $|k_\pm^2-\MW^2|/\MW^2$ and thus negligible in DPA. Therefore we can use the above assumptions and find that $\De$ vanishes in DPA
after averaging over the four points in the
$(k_+^2,k_-^2)$ plane that are related by the reflections \refeq{reflect}:
\beq\label{eq:Dcanc}
\De + \De\Big|_{K_+\to-K_+^*} + \De\Big|_{K_-\to-K_-^*} 
+ \De\Big|_{K_\pm\to-K_\pm^*} \sim 0.
\eeq
This explicitly confirms the results of \citere{Fa94}.
In particular, the non-factorizable corrections vanish in the limit 
$|k_\pm^2-\MW^2|\ll\GW\MW$, \ie for on-shell \PW~bosons.

The above considerations lead to the following simplified recipe for
the non-factorizable corrections to single invariant-mass distributions,
\ie  as long as at least one of the invariant masses $k_\pm^2$ is 
integrated out: the full factor $\De$ can be replaced according to
\beq
\label{recipe}
\De \;\to\; 
\frac{1}{2}\left(\Delta+\Delta\Big|_{K_- \leftrightarrow -K_-^*}\right) 
+ \frac{1}{2}\left(\Delta+\Delta\Big|_{K_+ \leftrightarrow -K_+^*}\right),
\eeq
where the first term on the r.h.s.\ is given in \refeq{kmaverage}, and
the second follows from the first by interchanging $K_+ \leftrightarrow
K_-$ and $s_{13} \leftrightarrow s_{24}$. Note that no double counting
is introduced, since the first (second) term does not contribute if
$k_+^2$ ($k_-^2$) is integrated out. 
In order to introduce the exact Coulomb singularity, one simply has to add the 
additional term of \refeq{demmcoul} 
to \refeq{kmaverage} and \refeq{recipe}.

\subsection{Non-factorizable corrections to related processes}

Since all non-factorizable corrections involving the initial $\Pep\Pem$ 
state cancel, the above results for the correction factor also apply 
to other \PW-pair production processes, such as 
$\gamma\gamma\to\PW\PW\to 4\,$fermions and $q\bar q\to\PW\PW\to 4\,$fermions.

The presented analytical results can also be carried over to
\PZ-pair-mediated four-fermion production, $\Pep\Pem\to\PZ\PZ\to
4\,$fermions.  In this case, $\Delta_{\ffp}$ yields the complete
non-factorizable correction, where all quantities such as $\MW$ and
$\Ga_\PW$ defined for the W~bosons are to be replaced by the ones for
the Z~bosons.  The fact that $Q_1=Q_2$ and $Q_3=Q_4$ has two important
consequences. Firstly, it implies the cancellation of mass
singularities contained in $\Delta_{\ffp}$ when all contributions are
summed as in \refeq{nfcorrfac}.  Secondly, it leads to the
antisymmetry of $\delta_\nf$ under each of the interchanges
$k_1\leftrightarrow k_2$ and $k_3\leftrightarrow k_4$. 
It is interesting to note that \refeq{nfcorrfac} with \refeq{eq:Delta}
are directly applicable, since 
$\De_{\mfp}$ and $\De_{\mmp}$ cancel in the sum of \refeq{nfcorrfac}. 
Therefore, a practical way to calculate the non-factorizable corrections
to $\Pep\Pem\to\PZ\PZ\to4\,$fermions consists in taking 
$\Delta_{\ffp}+\De_{\mfp}$ from \refeq{eq:DmfDPA} and \refeq{Fi}, and
setting $\De_{\mmp}$ to zero.

\section{Numerical results}
\label{senumres}

If not stated otherwise, we used the parameters
\beq\label{params}
\alpha ^{-1}=137.0359895, \quad \MZ=91.187\GeV, \quad
\MW=80.22\GeV, \quad \GW = 2.08\GeV,
\eeq
which coincide with those of \citere{Be97a},
for the numerical evaluation.

In order to exclude errors, we have written two
independent programs for the correction factor \refeq{nfcorrfac} and
compared all building blocks numerically.
These subroutines are implemented in the Monte Carlo generator
{\tt EXCALIBUR} \cite{Be94} as a correction factor to the three
(doubly-resonant) W-pair-production signal diagrams. In all numerical
results below, only these signal diagrams are included,
and no phase-space cuts have been applied.
We restrict our numerical analysis to the correction factor for purely
leptonic final states, 
$\Pep\Pem\to\PW\PW\to\nu_\ell \ell^+\ell^{\prime-}\nu_{\ell'}$. 
Results for other final states can 
easily be obtained from our formulae and computer programs. 

The results for the figures have been obtained from 50 million
phase-space points using the histogram 
routine of {\tt EXCALIBUR} with 40 bins for each figure.
For each entry in the tables,
10 million phase-space points were generated. 

\subsection{Comparison with existing results}

The non-factorizable photonic corrections have already been evaluated
by two groups. As was noted in the introduction, the
authors of \citeres{Be97a,Be97b} have not confirmed%
\footnote{While the result of \citeres{Be97a,Be97b} (as our result) for
the complete non-factorizable correction
is free of mass-singular logarithms, the result of \citere{Me96} 
contains logarithms of ratios of fermion masses. However, the authors of 
\citere{Me96} have informed us \cite{myprivcom} that the results of 
\citere{Me96} and \citeres{Be97a,Be97b} agree for equal fermion pairs in the 
final state.} 
the results of \citere{Me96}. Therefore, we first compare our 
findings with the results of these two groups.

\newcommand{\bk}{{\bf k}}
Melnikov and Yakovlev \cite{Me96} give the relative non-factorizable
corrections only to the completely differential cross section for the
process $\Pep\Pem\to\PW\PW\to\Pne\Pep\Pem\Pane$ as a function of the
invariant mass $M_+$ of the $\Pne\Pep$ system for all other
phase-space parameters fixed. All momenta of the final-state fermions
lie in a plane, and the momenta $\bk_2$ and $\bk_3$ of the final-state
positron and electron point into opposite directions. The angle
between the $\PWm$~boson and the positron is fixed to
$\theta_{\PWm\Pep}=150^\circ$, and the CM energy is chosen
as $\sqrt{s}=180\GeV$. The invariant mass of the $\Pem\Pane$ system
takes the values $M_-=78$ and $82\GeV$. The other parameters are
$\alpha=1/137$, $\MW=80\GeV$, and $\GW=2.0\GeV$.
In \reffi{fimy} we show our results for the non-factorizable photonic
corrections for this phase-space configuration.
\bfi
\centerline{
\setlength{\unitlength}{1cm}
\begin{picture}(7.5,7.8)
\put(0,0){\includegraphics{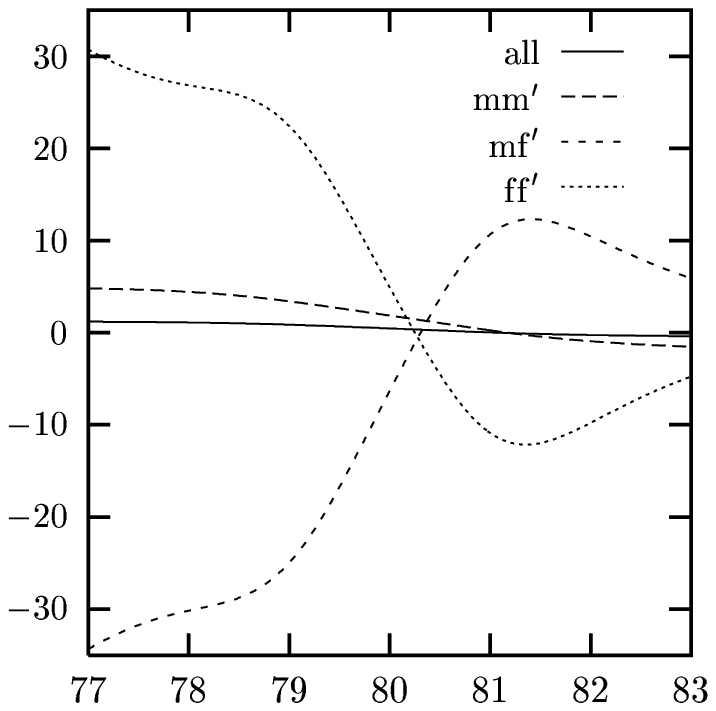}}
\put(0,5.0){\makebox(1,1)[c]{$\de_\nf/\%$}}
\put(4.5,-0.3){\makebox(1,1)[cc]{{$M_+/{\rm GeV}$}}}
\put(4.6,2){$M_-=78\GeV$}
\end{picture}
\begin{picture}(7.5,7.8)
\put(0,0){\includegraphics{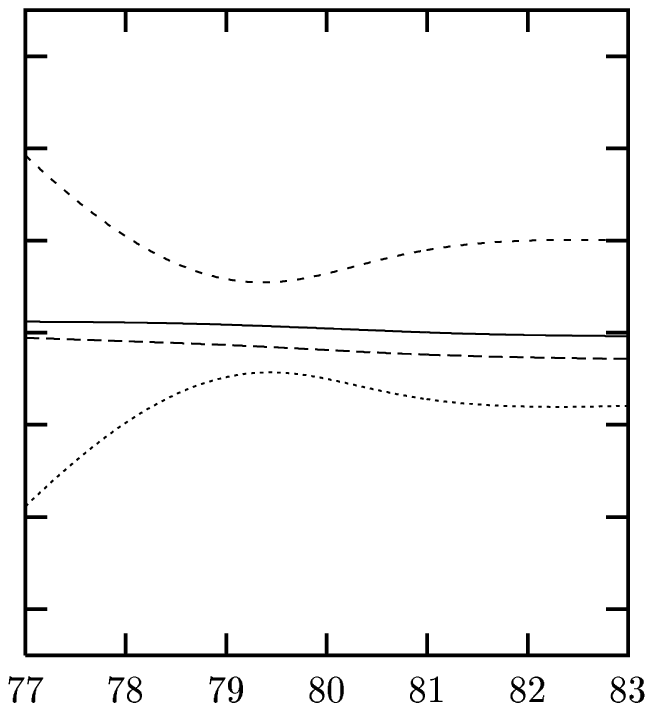}}
\put(3.5,-0.3){\makebox(1,1)[cc]{{$M_+/{\rm GeV}$}}}
\put(4,2){$M_-=82\GeV$}
\end{picture}
}
\caption{Relative non-factorizable correction factor to the
  differential cross section for the phase-space configuration
  given in the text.
  The curves labelled \mmp, \mfp, and \ffp~correspond
  to the curves A, B, and C of \protect\citere{Me96}, respectively.}
\label{fimy}
\efi
The intermediate--intermediate (\mmp)
corrections agree reasonably with those of \citere{Me96}, 
but the other curves differ qualitatively and
quantitatively%
\footnote{\samepage{Although not stated in \citere{Me96}, mass-singular parts
have been dropped there in the numerical evaluation \cite{myprivcom}
rendering a thorough comparison of the (\mfp) and (\ffp) parts
impossible. Comparing the sum of all three contributions, i.e.\ the
complete non-factorizable correction factor, our result differs from 
the sum read off from \citere{Me96}.}}.
We mention that \reffi{fimy} has been reproduced \cite{bbcprivcom} by the
authors of \citeres{Be97a,Be97b} within the expected level of accuracy.
While the individual contributions shown in
\reffi{fimy} are at the level of 10\% owing to mass singularities, the
sum, which is free of mass singularities, is below 1.2\%.

Beenakker et al.\ \cite{Be97a} have evaluated the relative non-factorizable
corrections to the distributions 
$\rd\sigma/\rd M_+\rd M_-$, $\rd\sigma/\rd M_+$, and
$\rd\sigma/\rd M_{\mathrm{av}}$, where $M_{\mathrm{av}}=(M_-+M_+)/2$.
Our corresponding results for the set of parameters given in \refeq{params}
are shown in \reffi{fi3bbc} for the single
invariant-mass distributions and in 
\refta{tab1bbc} for the double invariant-mass distribution. 
\bfi
\centerline{
\setlength{\unitlength}{1cm}
\begin{picture}(10,7.8)
\put(0,0){\includegraphics{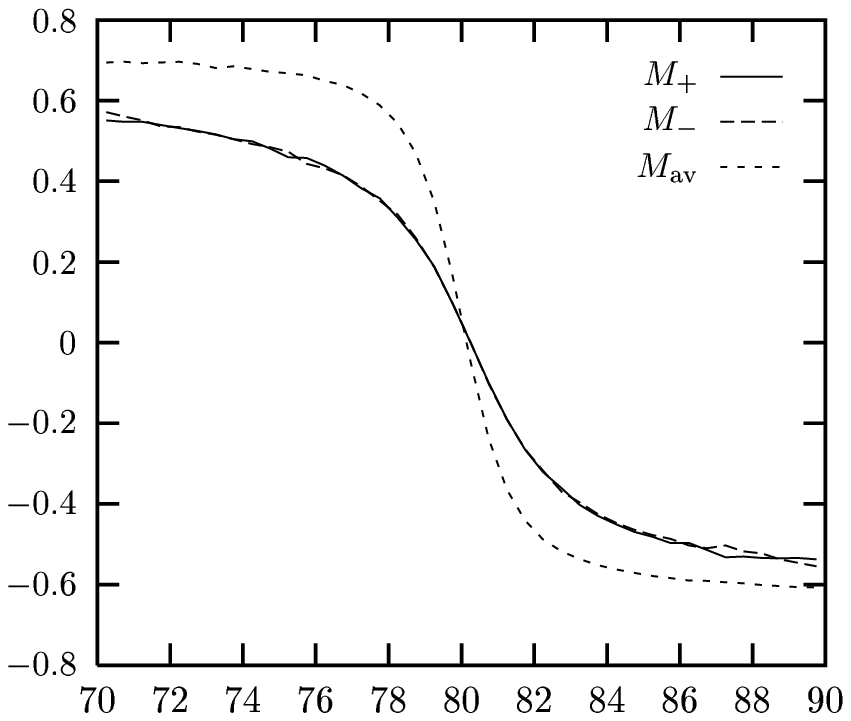}}
\put(-0.5,5.0){\makebox(1,1)[c]{$\de_\nf/\%$}}
\put(4.7,-0.3){\makebox(1,1)[cc]{{$M/{\rm GeV}$}}}
\end{picture}
}
\caption[]{Relative non-factorizable corrections 
  to the single invariant-mass distributions
  $\protect\rd\sigma/\protect\rd M_\pm$ and
  $\protect\rd\sigma/\protect\rd M_{\protect\mathrm{av}}$ for the
    CM energy $\sqrt{s}=184\protect\GeV$.}
\label{fi3bbc}
\efi
\newcommand{\mspc}{\phantom{-}}
\begin{table}
\renewcommand{\arraycolsep}{2ex}
\[
  \begin{array}{||c||c|c|c|c|c|c|c||}    \hline \hline
    \raisebox{-3mm}{$\Delta_+$}
    & \multicolumn{7}{|c||}{\raisebox{-1mm}{$\Delta_-$}}\\
    \cline{2-8}
      & -1    & -1/2  & -1/4  &  0    & 1/4   & 1/2   & 1     \\
    \hline  \hline
 -1  &\mspc0.81 &\mspc0.64 &\mspc0.52 &\mspc0.38 &\mspc0.22 &\mspc0.07 & -0.16\\
-1/2 &\mspc0.64 &\mspc0.52 &\mspc0.40 &\mspc0.24 &\mspc0.08 &    -0.07 & -0.25\\
-1/4 &\mspc0.52 &\mspc0.40 &\mspc0.28 &\mspc0.13 &    -0.02 &    -0.15 & -0.31\\
  0  &\mspc0.38 &\mspc0.24 &\mspc0.13 &\mspc0.00 &    -0.13 &    -0.24 & -0.37\\
 1/2 &\mspc0.22 &\mspc0.08 &    -0.02 &    -0.13 &    -0.24 &    -0.32 & -0.43\\
 1/4 &\mspc0.07 &    -0.07 &    -0.15 &    -0.24 &    -0.32 &    -0.39 & -0.48\\
  1  &    -0.16 &    -0.25 &    -0.31 &    -0.37 &    -0.43 &    -0.48 & -0.54\\
    \hline \hline
  \end{array}
\]
\caption[]{Relative non-factorizable corrections 
           in per cent to the double invariant-mass distribution
           $\rd\sigma/\rd M_+\rd M_-$ for the CM energy 
           $\sqrt{s}=184\GeV$ and various values of $M_{\pm}$
           specified in terms of their
           distance from $\MW$ in units of $\GW$,
           \ie $\Delta_{\pm} = (M_{\pm}-\MW)/\GW$.}
\label{tab1bbc}
\end{table}%
The deviation between the distributions $\rd\sigma/\rd M_+$ and
$\rd\sigma/\rd M_-$ in \reffi{fi3bbc}, which should be identical,
gives an indication on the Monte Carlo error of our calculation.  The
single and double invariant-mass distributions agree very well with
those of \citere{Be97a}. 
The worst agreement is found for small invariant masses and amounts to
0.03\%. In fact, the agreement is better than expected,
in view of the fact that our results differ from those of
\citere{Be97a} by non-doubly resonant corrections. 
In the numerical evaluations of \citere{Be97a} the phase space and the
Born matrix element are taken entirely on shell \cite{bbcprivcom}.
Moreover, the scalar integrals are parametrized by scalar invariants
different from ours, leading to differences of the order
of $|k_\pm^2-\MW^2|/\MW^2$.

In \citere{Be97a}, additionally, the decay-angular distribution 
$\rd\sigma/\rd M_-\rd M_+\rd\cos\theta_{\PWp\Pep}$ has been considered,
where $\theta_{\PWp\Pep}$ is the decay angle between $\bk_+$ and $\bk_2$ 
in the laboratory system. 
Our results for this angular distribution are shown in \reffi{fi2bbc}.
\bfi
\centerline{
\setlength{\unitlength}{1cm}
\begin{picture}(10,7.8)
\put(0,0){\includegraphics{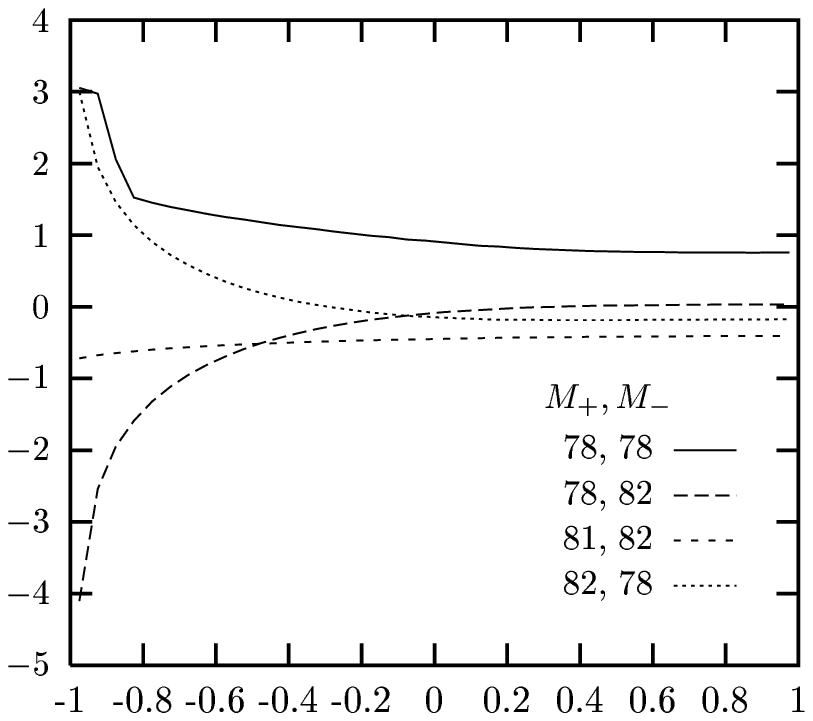}}
\put(-0.5,5.0){\makebox(1,1)[c]{$\de_\nf/\%$}}
\put(4.7,-0.3){\makebox(1,1)[cc]{{$\cos\theta_{\PWp\Pep}$}}}
\end{picture}
}
\caption[]{Relative non-factorizable corrections to the decay-angular
           distribution $\rd\sigma/\rd M_-\rd M_+\rd\cos\theta_{\PWp\Pep}$
           for fixed values of the invariant masses $M_{\pm}$ 
           and the CM energy $\sqrt{s}=184\protect\GeV$.}
\label{fi2bbc}
\efi
The cross section is small for $\cos\theta_{\PWp\Pep}\sim-1$, where
the corrections are largest. 
Unfortunately the corresponding figure in \citere{Be97a} is not correct 
\cite{bbcprivcom}. The authors of \citere{Be97a} have provided a 
corrected figure, which agrees reasonably well with \reffi{fi2bbc}, but 
does not show the kinks in the curve for $M_\pm=78\GeV$. 
The kinks are due to a logarithmic Landau singularity in the 
4-point functions.
If one employs the on-shell parametrization of phase space, 
as in \citere{Be97a}, the Landau
singularities appear at the boundary of phase space. Although no kinks
appear in the physical phase space in this case, the Landau
singularities still give rise to large corrections for
$\cos\theta_{\PWp\Pep}\sim-1$. 
Since the kinks appear in a region where the cross section is
small, they are not relevant for phenomenology.
The issue of the kinks is further discussed in \refse{se:ambig}.

\subsection{Further results}

In \reffi{energyplot} we show the non-factorizable corrections to the
single invariant-mass distribution $\rd\sigma/\rd M_+$ for various
CM energies.  While the corrections reach up to 1.3\% for $\sqrt{s}=172\GeV$,
they decrease with increasing energy and are less than 0.04\%
for $\sqrt{s}=300\GeV$. 
\bfi
\centerline{
\setlength{\unitlength}{1cm}
\begin{picture}(10,7.8)
\put(0,0){\includegraphics{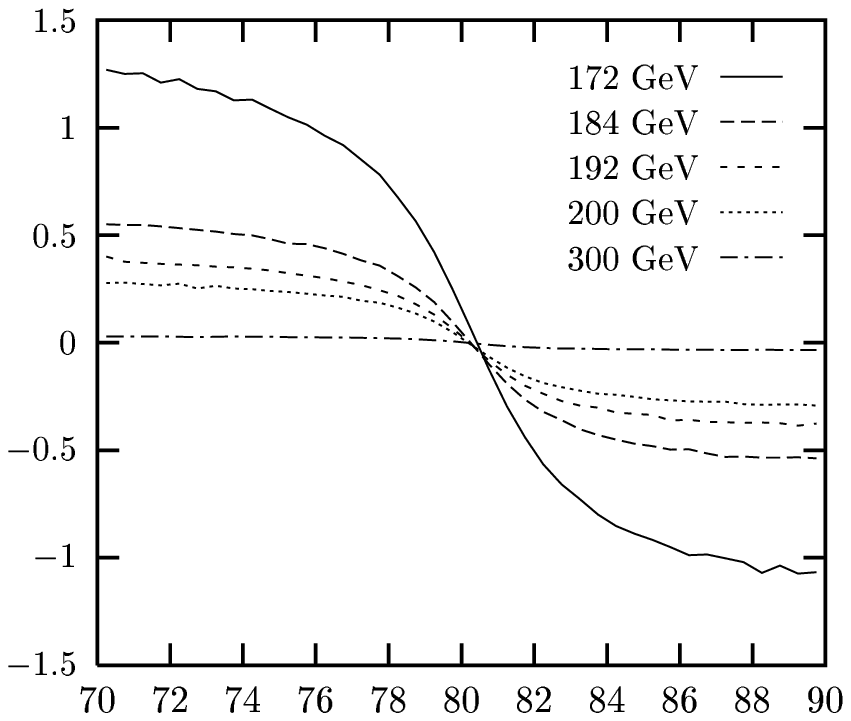}}
\put(-0.5,5.0){\makebox(1,1)[c]{$\de_\nf/\%$}}
\put(4.7,-0.3){\makebox(1,1)[cc]{{$M_+/{\rm GeV}$}}}
\end{picture}
}
\caption[]{Relative non-factorizable corrections 
  to the single invariant-mass distribution 
  $\protect\rd\sigma/\protect\rd M_+$ for
  various CM energies.}
\label{energyplot}
\efi
Note that the shape of the corrections is exactly what is naively
expected. If a photon is emitted in the final state, the invariant mass 
of the fermion pair is smaller than the invariant mass of the resonant
\PW~boson, which is given by the invariant mass of the fermion pair plus 
photon. Since we calculate the corrections as a function of the invariant
masses of the fermion pairs, the cross section tends to increase
for small invariant masses and decrease for large invariant masses.

The non-factorizable corrections distort the invariant-mass
distribution and thus lead to a shift in the \PW-boson mass determined
from the direct reconstruction of the decay products with respect to the
actual \PW-boson mass. This shift can be estimated by the
displacement of the maximum of the single-invariant-mass distribution
caused by the corrections shown in \reffi{energyplot}. 
To this end, we determine
the slope of the corrections for $M_+=\MW$, multiply this linearized
correction to a simple Breit-Wigner factor, and determine the shift 
$\Delta M_+$ of the maximum. The smallness of the correction allows us
to evaluate $\Delta M_+$ in linear approximation, leading to the simple
formula
\beq
\Delta M_+ = 
\biggl(\frac{\rd\delta_{\nf}}{\rd M_+}\biggr)\bigg|_{M_+=\MW} 
\frac{\GW^2}{8}.  
\eeq
Extracting the slope from our numerical results we obtain the mass shifts 
shown in \refta{ta:shifts}.
\begin{table}
$$
\renewcommand{\arraycolsep}{2ex}
\begin{array}{|c|r|r|r|r|r|}
\hline
\sqrt{s}/\mathrm{GeV}    &   172  &    184   &  192  &  200  &   300 \\
\hline
\Delta M_+/\mathrm{MeV}  &  -2.0  &   -1.1   & -0.8  & -0.6  &  -0.09
\\
\hline
\end{array}
$$
\caption{Shift of the maximum of the single invariant-mass
  distributions $\rd\si/\rd M_+$ induced by the non-factorizable
  corrections at various CM energies.}
\label{ta:shifts}
\end{table}

In \reffis{phiplot} and \ref{thetaepem} we show the 
effect of the non-factorizable corrections on various angular distributions. 
\bfi
\centerline{
\setlength{\unitlength}{1cm}
\begin{picture}(10,7.8)
\put(0,0){\includegraphics{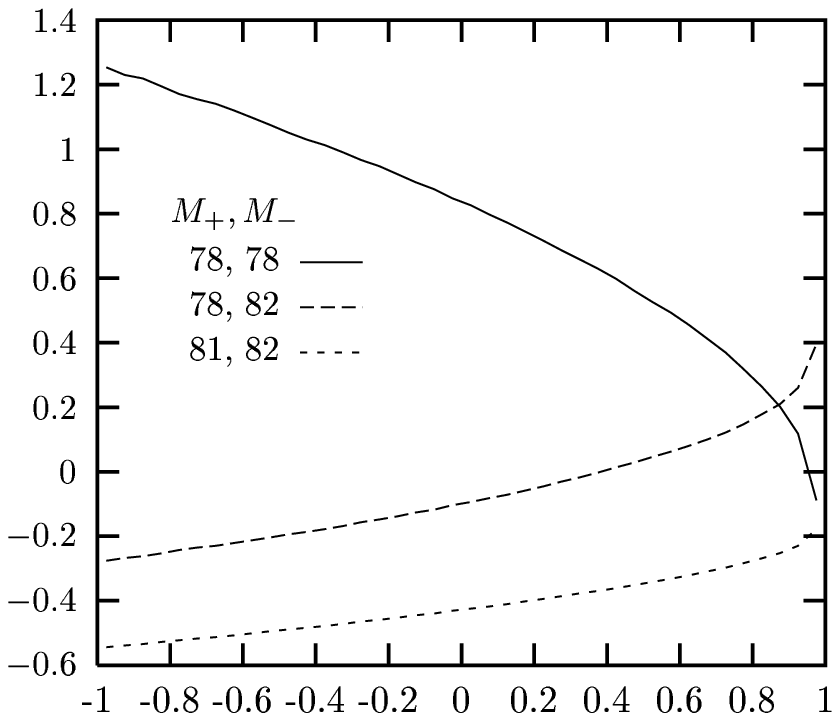}}
\put(-0.5,5.0){\makebox(1,1)[c]{$\de_\nf/\%$}}
\put(4.7,-0.3){\makebox(1,1)[cc]{{$\cos\phi$}}}
\end{picture}
}
\caption[]{Relative non-factorizable corrections to the angular
             distribution $\protect\rd\sigma/\protect\rd M_-\rd M_+\rd\cos\phi$
             for fixed values of the invariant masses $M_{\pm}$ 
             and the CM energy $\sqrt{s}=184\protect\GeV$.}
\label{phiplot}
\efi
\bfi
\centerline{
\setlength{\unitlength}{1cm}
\begin{picture}(10,7.8)
\put(0,0){\includegraphics{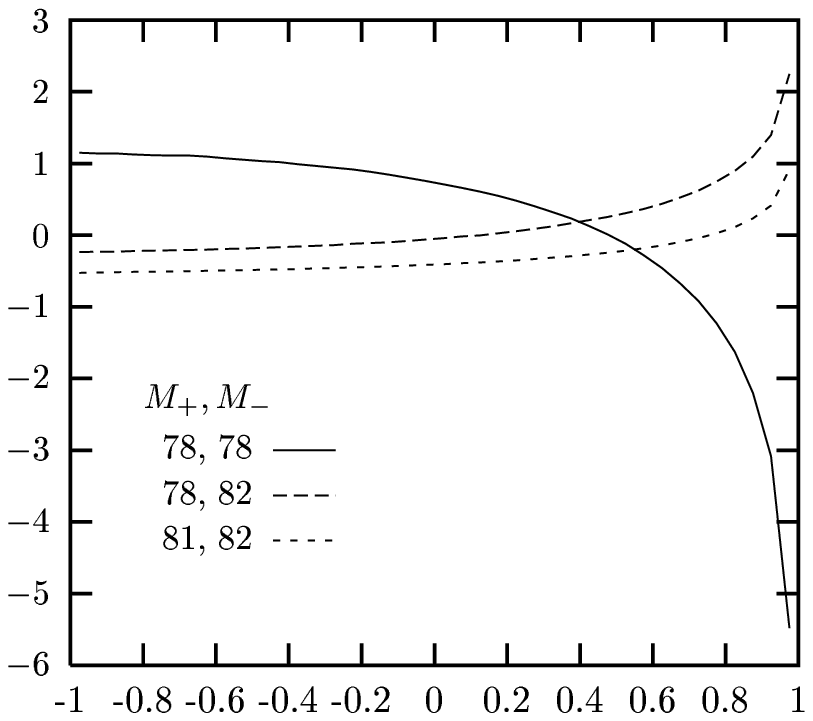}}
\put(-0.5,5.0){\makebox(1,1)[c]{$\de_\nf/\%$}}
\put(4.7,-0.3){\makebox(1,1)[cc]{{$\cos\theta_{\Pep\Pem}$}}}
\end{picture}
}
\caption[]{Relative non-factorizable corrections to the angular distribution 
             $\protect\rd\sigma/\protect\rd M_-\rd M_+\rd\cos\theta_{\Pep\Pem}$
             for fixed values of the invariant masses $M_{\pm}$ 
             and a CM energy $\sqrt{s}=184\protect\GeV$.}
\label{thetaepem}
\efi
Since the non-factorizable corrections are independent of the production angle
of the \PW~bosons, it suffices to consider distributions
involving the angles of the final-state fermions. 
We define all angles in the laboratory system,
which is the CM system of the production process. 
The distribution over the angle $\phi$ between the two planes spanned
by the momenta of the two fermion pairs in which the \PW~bosons decay, \ie
\beq
\cos\phi = \frac{(\bk_1\times \bk_2)(\bk_3\times \bk_4)}
{|\bk_1\times \bk_2||\bk_3\times \bk_4|},
\eeq
is presented in \reffi{phiplot}. 
The corrections are of the order of 1\% or less. Like
the $\phi$ distribution, the distribution over the angle between positron 
and electron $\theta_{\Pep\Pem}$ (\reffi{thetaepem}) is symmetric under the
interchange of $M_+$ and $M_-$. 
As for the $\theta_{\Pep\PWp}$ distribution (\reffi{fi2bbc}), 
the corrections reach several per
cent in the region where the cross section is small.

The distribution in the electron energy $E_\Pem$ is considered in \reffi{Eem}.
\bfi
\centerline{
\setlength{\unitlength}{1cm}
\begin{picture}(10,7.8)
\put(0,0){\includegraphics{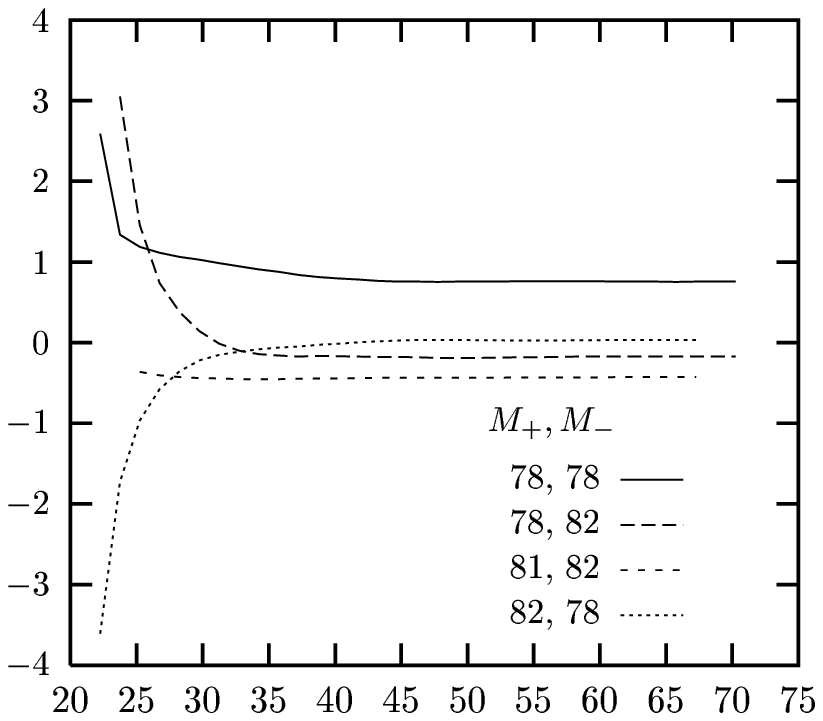}}
\put(-0.5,5.0){\makebox(1,1)[c]{$\de_\nf/\%$}}
\put(4.7,-0.3){\makebox(1,1)[cc]{{$E_\Pem/\GeV$}}}
\end{picture}
}
\caption[]{Relative non-factorizable corrections to the electron-energy
distribution 
$\protect\rd\sigma/\protect\rd M_-\rd M_+\rd E_\Pem$
for fixed values of the invariant masses $M_{\pm}$ 
and a CM energy $\sqrt{s}=184\protect\GeV$.}
\label{Eem}
\efi
The corrections are typically of the order of 1\%. Again the 
corrections become large where the cross section is small.

In \refse{se43} we have introduced a correction term that includes the
full off-shell Cou\-lomb singularity. The results for the
non-factorizable corrections with this improvement
are compared with those of the pure DPA in \reffi{bbcfig3coul}.
\bfi
\centerline{
\setlength{\unitlength}{1cm}
\begin{picture}(10,7.8)
\put(0,0){\includegraphics{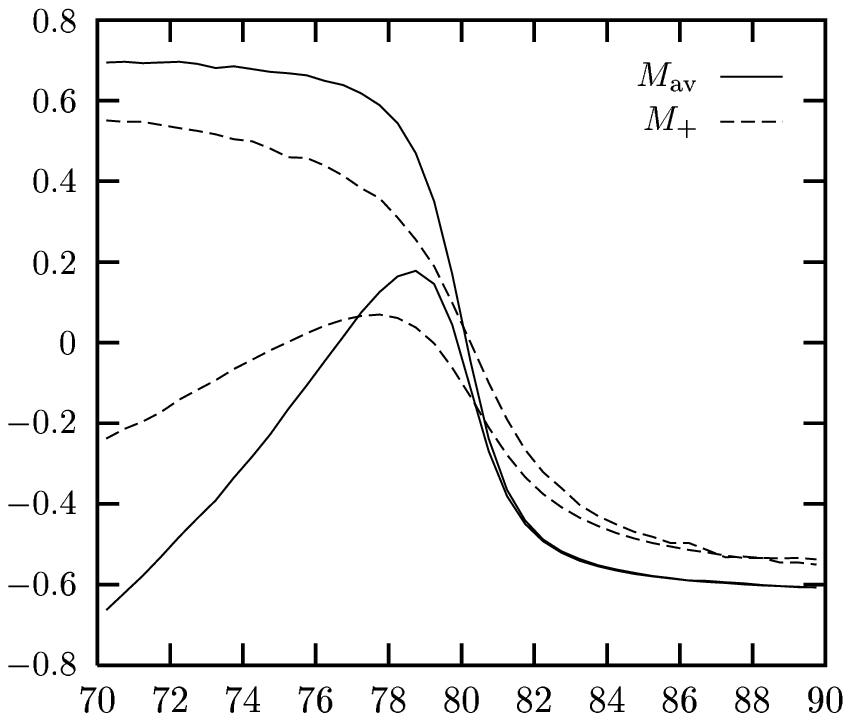}}
\put(-0.5,5.0){\makebox(1,1)[c]{$\de_\nf/\%$}}
\put(4.7,-0.3){\makebox(1,1)[cc]{{$M/{\rm GeV}$}}}
\end{picture}
}
\caption[]{Relative non-factorizable corrections 
  to the single invariant-mass distributions
  $\protect\rd\sigma/\protect\rd M_+$ and
  $\protect\rd\sigma/\protect\rd M_{\protect\mathrm{av}}$ for a CM
    energy $\sqrt{s}=184\protect\GeV$ with (lower curves) and without 
    (upper curves) improved Coulomb-singularity treatment.}
\label{bbcfig3coul}
\efi
\begin{table}
\renewcommand{\arraycolsep}{2ex}
\[
  \begin{array}{||c||c|c|c|c|c|c|c||}    \hline \hline
    \raisebox{-3mm}{$\Delta_+$}
    & \multicolumn{7}{|c||}{\raisebox{-1mm}{$\Delta_-$}}\\
    \cline{2-8}
         & -1    & -1/2  & -1/4  &  0    & 1/4   & 1/2   & 1     \\
    \hline  \hline
  -1   & \mspc0.39 &   \mspc 0.31 &   \mspc 0.23 &   \mspc 0.15 &   \mspc 0.05 &   -0.04 &   -0.20\\
 -1/2  &\mspc0.31 &\mspc0.28 &\mspc0.20 &\mspc0.10 &   -0.02 & -0.13 & -0.27\\
 -1/4  &\mspc0.23 &\mspc0.20 &\mspc0.13 &\mspc0.03 &   -0.09 & -0.19 & -0.33\\
   0   &\mspc0.15 &\mspc0.10 &\mspc0.03 &   -0.08  &   -0.18 & -0.27 & -0.38\\
  1/2  &\mspc0.05 &   -0.02  &   -0.09  &   -0.18  &   -0.27 & -0.35 & -0.44\\
  1/4  &   -0.04  &   -0.13  &   -0.19  &   -0.27  &   -0.35 & -0.41 & -0.49\\
   1   &   -0.20  &   -0.27  &   -0.33  &   -0.38  &   -0.44 & -0.49 & -0.54\\
    \hline \hline
  \end{array}
\]
\caption[]{Same as in \refta{tab1bbc} but with improved Coulomb-singularity 
treatment.}
\label{tab1bbcwc}
\end{table}%
For $\sqrt{s}=184\GeV$
the additional terms shift the non-factorizable corrections by up to 1.4\%
for $\protect\rd\sigma/\protect\rd M_{\protect\mathrm{av}}$ and
by up to 0.8\% for $\protect\rd\sigma/\protect\rd M_+$
for small invariant masses, whereas for large invariant masses
there is practically no effect.
The difference originates essentially from the differences between $1/\betap$ 
and $1/\betaW$ in \refeq{demmcoul}. For large invariant masses, 
the explicit logarithms in \refeq{demmcoul} are small, \ie
the Coulomb singularity correction is minuscule, and
this difference practically makes no effect. For small invariant
masses, the logarithms are approximately $\ri\pi$ and the
difference causes the effect seen in \reffi{bbcfig3coul}. 
In \refta{tab1bbcwc} we show the non-factorizable corrections to the
double invariant-mass distribution, as in \refta{tab1bbc}, but now with
the improved Coulomb-singularity treatment.
We find a difference of up to half a per cent for small invariant masses 
but no effect for large ones.
We mention that the difference between the entries in \reftas{tab1bbcwc} 
and \ref{tab1bbc} is directly given by the contribution \refeq{demmcoul}
to $\Delta_{\mathrm{mm}'}$, without any influence of the phase-space
integration.

\subsection{Discussion of intrinsic ambiguities}
\label{se:ambig}

In the results presented so far, all scalar integrals were
parametrized by $s$, $s_{23}$, $s_{13}$, $s_{24}$, and $k_\pm^2$
(parametrization 1). In DPA, however, the parameters of the
scalar integrals are only fixed up to terms of order $k_\pm^2-\MW^2$. 
We can for example parametrize the scalar integrals in terms of $s$,
$s_{23}$, $s_{123}$, $s_{234}$, and $k_\pm^2$ (parametrization 2) instead. 
As a third parametrization, we fix all scalar invariants except for
$k_\pm^2$ by their on-shell values, corresponding exactly to the
approach of \citere{Be97a}. The results of these three parametrizations 
differ by non-doubly-resonant corrections. 

The difference between parametrizations 1 and 2 is illustrated
in \reffi{vglparamM1} for the single invariant-mass distribution.
\bfi
\centerline{
\setlength{\unitlength}{1cm}
\begin{picture}(10,7.8)
\put(0,0){\includegraphics{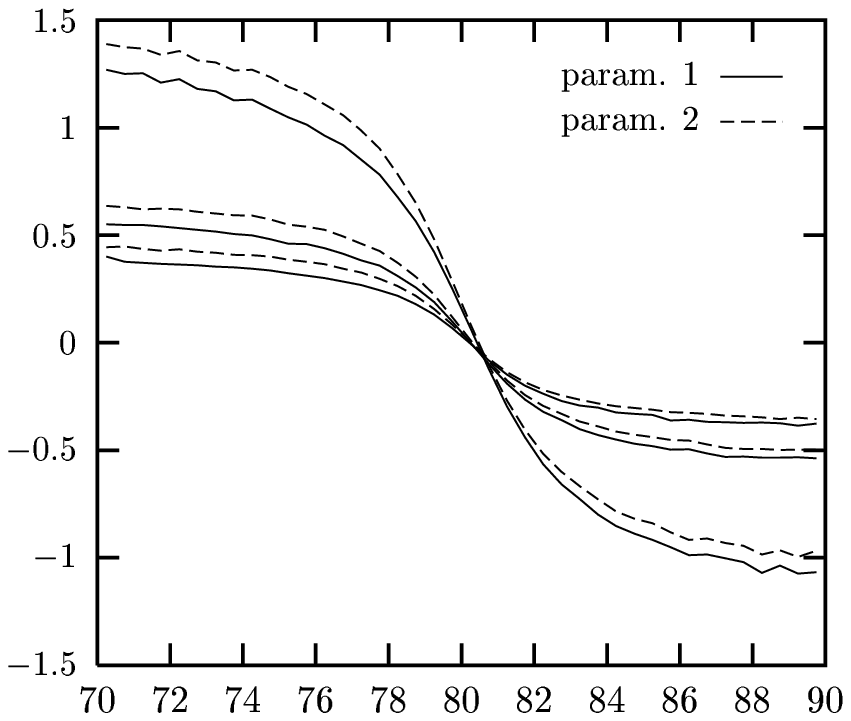}}
\put(-0.5,5.0){\makebox(1,1)[c]{$\de_\nf/\%$}}
\put(4.7,-0.3){\makebox(1,1)[cc]{{$M_+/\protect\mathrm{GeV}$}}}
\put(6.9,4.0){$\scriptstyle\sqrt{s}=192\GeV$}
\put(6.9,2.8){$\scriptstyle\phantom{\sqrt{s}=}184\GeV$}
\put(6.9,1.7){$\scriptstyle\phantom{\sqrt{s}=}172\GeV$}
\end{picture}
}
\caption{Relative non-factorizable corrections 
  to the single invariant-mass distribution
  $\protect\rd\sigma/\protect\rd M_+$ for 
  the CM energies 
  $172$, $184$, and $192\GeV$ and two different parametrizations.}
\label{vglparamM1}
\efi
\bfi
\centerline{
\setlength{\unitlength}{1cm}
\begin{picture}(14.5,7.8)
\put(0.5,0){\includegraphics{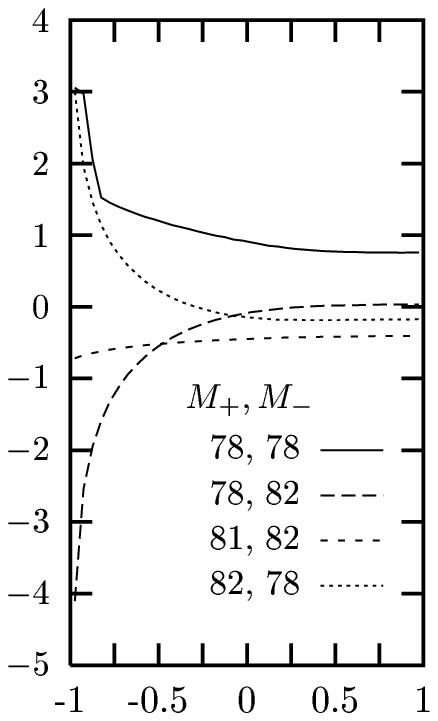}}
\put(4.8,0){\includegraphics{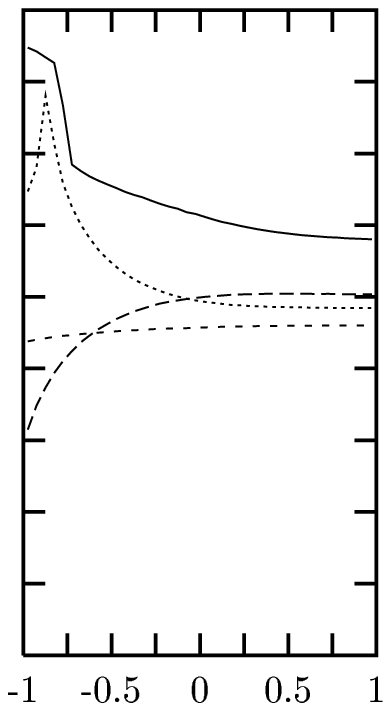}}
\put(9.1,0){\includegraphics{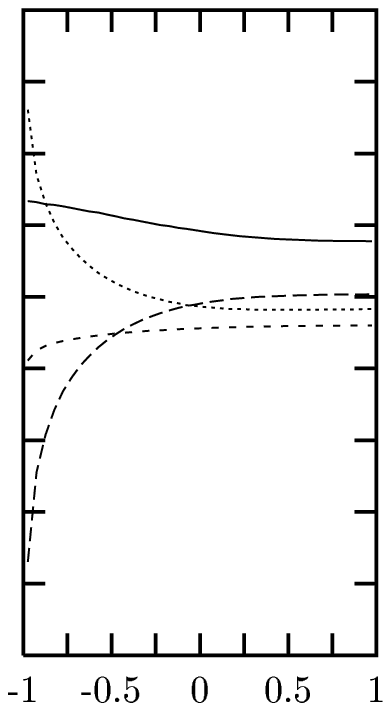}}
\put(-0.2,5.0){\makebox(1,1)[c]{$\de_\nf/\%$}}
\put(7.0,-0.3){\makebox(1,1)[cc]{{$\cos\theta_{\PWp\Pep}$}}}
\put( 2.6,7.0){param.~1}
\put( 6.9,7.0){param.~2}
\put(11.2,7.0){param.~3}
\end{picture}
}
\caption[]{Relative non-factorizable corrections
             to the decay-angular
             distribution $\rd\sigma/\rd M_-\rd M_+\rd\cos\theta_{\PWp\Pep}$
             for fixed values of the invariant masses $M_{\pm}$ 
             and a CM energy $\sqrt{s}=184\protect\GeV$
             using three different parametrizations, as specified in
             the text.}
\label{vglparamtheta1}
\efi
The difference amounts to $\sim 0.1\%$.
Note that for an invariant mass $M_+ = 70\GeV$ we have 
$\al|\MW^2-k_+^2|/\MW^2\sim 0.002$ and would thus expect absolute
changes in the non-factorizable corrections at this level.  

For the non-factorizable corrections to the angular distributions,
uncertainties of the same order are to be expected. The only exceptions
are the distributions over the decay angles $\theta_{\PWp\Pep}$ and
$\theta_{\PWm\Pem}$. Let us explain this for $\theta_{\PWp\Pep}$ in more
detail: the non-factorizable correction contains the term
$2\pi\ri\ln[1+x_\PW\MW^2/(s_{13}+s_{23})]$, which can be evaluated by taking
$(s_{13}+s_{23})$ directly or $(s_{123}-\MW^2)$ as input. This
parametrization ambiguity can lead to larger uncertainties, because the
above logarithm can become singular, and the location of this Landau
singularity is shifted by the ambiguity. Since there is a one-to-one
correspondence between $s_{123}$ and $\theta_{\PWp\Pep}$ for fixed $s$
and $k_+^2$, this logarithmic singularity is washed out if the angular 
integration over $\theta_{\PWp\Pep}$ is performed, but appears as a kink 
structure in the angular distribution over this angle.
Figure~\ref{vglparamtheta1} shows the non-factorizable corrections to
this angular distribution for the three parametrizations.
For $M_+ = 78\GeV$ we still have $\alpha|k_+^2-\MW^2|/\MW^2\sim 0.0004$. 
Apart from the regions where the Landau singularities appear, this is
indeed of the order of the differences between the three parametrizations.
When considering the Landau singularities, one should realize that the
parametrization ambiguity of the locations of the singularities is not
suppressed by a factor $\alpha$, \ie different parametrizations shift
the locations at the level of $|k_+^2-\MW^2|/\MW^2\sim 0.05$.
However, the impact of the corresponding kinks on observables is again
suppressed with $\alpha|k_+^2-\MW^2|/\MW^2$ if the angles are integrated over, 
since the singularities can only appear near the boundary of phase space 
and disappear from phase space exactly on resonance. 
Since the cross section is small where the Landau singularity appears,
the effect is phenomenologically irrelevant%
\footnote{ \samepage{
Although it would be possible to remove all Landau singularities from
the physical phase space in DPA by choosing an appropriate
parametrization (e.g.\ parametrization 3), we have
kept these structures, because we view this kind of
ambiguity as unavoidable in DPA. Only a complete off-shell calculation
in $\O(\alpha)$ can resolve this uncertainty. We have checked that 
the Landau singularities are definitely present in the scalar integrals
appearing in the exact off-shell calculation and we do not see a reason
why these should drop out in the complete off-shell $\O(\alpha)$ calculation. 
In any case, the kink structures will be smeared
out by the finite decay width of the W~boson.}}.

\section{Conclusion}

A reasonable approach for the evaluation of the $\Oa$ corrections to 
\PW-pair-mediated four-fermion production
consists in an expansion around the poles of the
\PW~propagators. For most of the corrections, the leading terms in this
expansion, the doubly-resonant corrections, are sufficient. Combining
the results for on-shell \PW-pair production and decay, one obtains the 
factorizable doubly-resonant corrections,
which account for the largest part of the corrections.
The remaining doubly-resonant corrections comprise
the effects of virtual soft-photon exchange and real
soft-photon interference where \PW-pair production and \PW~decay are not
independent. Those corrections are called non-factorizable.

We have studied the non-factorizable doubly-resonant corrections 
to \PW-pair production analytically as well as numerically.
By defining these corrections in such a way that
the complete doubly-resonant corrections are obtained by adding the
factorizable corrections, 
the non-factorizable corrections become manifestly gauge-independent. 
We have modified the double-pole approximation
in such a way that the non-factorizable corrections also include the
correct off-shell Coulomb singularity, which is 
the most important effect of the instability of the \PW~bosons.

It turns out that the actual form of the real non-factorizable corrections
depends on the parametrization of phase space, \ie on whether the
photon momentum is included in the definition of the invariant masses
of the resonant \PW~bosons or not.  This ambiguity is due to the fact that 
not only extremely soft photons but also photons with energies of the order
of the \PW~width contribute to the doubly-resonant corrections.
For the parametrization where the invariant masses of the final-state
fermion pairs are fixed when integrating over the photon momentum, we
have listed all relevant analytical results.
We have sketched the calculation of the real
bremsstrahlung integrals and the reduction of the real 5-point functions.
The final analytical results are presented as transparent as possible so
that they can be easily implemented in Monte Carlo generators. 
Our correction factor for the non-factorizable corrections agrees
analytically with the one of \citeres{Be97a,Be97b} in double-pole
approximation.

We have numerically discussed various distributions and, in particular,
compared to the results of \citeres{Me96,Be97a} as far as possible.
Our numerical results for the correction factor 
differ from those of \citere{Me96},
but agree with those of \citere{Be97a} at the level of the expected
intrinsic uncertainty of the double-pole approach. These small
deviations are due to different embeddings of the non-factorizable
correction in the off-shell phase space, which are equivalent in
double-pole approximation.

For the adopted phase-space parametrization 
the non-factorizable doubly-resonant corrections are independent of
the \PW-production angle, because all non-factorizable virtual and real
soft-photonic effects that are related to the initial state cancel.
For this reason, the presented correction factor
for the non-factorizable doubly-resonant corrections
is directly applicable to other reactions with the same final state,
such as $\ga\ga\to\PW\PW\to 4\,$fermions and 
$q\bar q\to\PW\PW\to 4\,$fermions.
The non-factorizable corrections to \PZ-pair-mediated
four-fermion production, $\Pep\Pem\to\PZ\PZ\to 4\,$fermions, can also be
read off from our analytical results. 

The non-factorizable corrections are non-vanishing only 
for distributions in the invariant masses of the \PW~bosons
and vanish after integration over both invariant masses.
Therefore, they do not modify the usual 
genuine differential distributions that are used for
the investigation of the triple gauge couplings.
For invariant-mass distributions, 
the non-factorizable doubly-resonant corrections
in the LEP2 energy region are small 
with respect to the experimental accuracy
at LEP2, but should become relevant for possible future linear
colliders with higher luminosity.
At high energies the non-factorizable corrections are negligible.

\section*{Acknowledgements}

\begin{sloppypar}
We thank R.~Pittau for help in implementing the corrections in EXCALIBUR.
We are grateful to W.~Beenakker, F.A.~Berends, A.P.~Chapovsky, 
K.~Melnikov, R.~Pittau, D.~Wackeroth and O.~Yakovlev for useful discussions.
Finally, we thank A.P.~Chapovsky for his effort to perform a detailed
comparison with our numerical results for the observables discussed in
\citere{Be97a}.
\end{sloppypar}

\appendix

\section*{Appendix}

\section{Useful definitions}
\label{prelim}

In the main text, we have already used the following short-hand expressions,
\beqar\label{abbrev}
\beta_\PW &=& \sqrt{1-\frac{4\MW^2}{s}+\ri\epsilon},
\nl
x_\PW &=& \frac{\beta_\PW-1}{\beta_\PW+1},
\nl
\beta &=&  \sqrt{1-\frac{4M^2}{s}},\nl
\bar\beta &=& \frac{\sqrt{\lambda(s,k_+^2,k_-^2)}}{s},\nl
\kappa_\PW &=& \sqrt{ \lambda(\MW^4,s_{13}s_{24},s_{14}s_{23})
        -\ri\epsilon},\nl
\Delta_M &=& \frac{|k_+^2-k_-^2|}{s},
\eeqar
where 
\beq
\lambda(x,y,z) = x^2+y^2+z^2-2xy-2xz-2yz.
\eeq
The evaluation of one-loop $n$-point functions naturally leads
to the usual dilogarithm,
\beq
\Li(z) = -\int_0^z\,\frac{\rd t}{t}\,\ln(1-t),
\qquad |\arc(1-z)|<\pi, 
\eeq
and its analytically continued form
\beqar
\cLi(x,y) &=& \Li(1-xy)+[\ln(xy)-\ln(x)-\ln(y)]\ln(1-xy),
\nl && 
|\arc(x)|,|\arc(y)|<\pi. 
\eeqar

\section{Calculation of the real bremsstrahlung integrals}
\label{appbrcal}

In this appendix we describe a general method for evaluating the
bremsstrahlung 3- and 4-point integrals defined in \refeq{realCDE0}. 
We make use of the generalized Feynman-parameter representation \cite{Fpar}
\begin{equation}\label{eq:FeynPar}
\frac{1}{\prod_{i=1}^n N'_i} =
\Gamma(n) \int_0^\infty \rd x_1 \cdots \rd x_n  
\frac{\delta (1- \sum_{i=1}^n \alpha_i x_i)}{(\sum_{i=1}^n N'_i x_i)^n},
\end{equation} 
where the real variables $\alpha_i\ge0$ are arbitrary, but not all 
equal to zero simultaneously. The sum $\sum_{i=1}^n N'_i x_i$ must be 
non-zero over the entire integration domain.

We first consider the 3-point integral $\Cbr_0$. We only need the difference 
between the general IR-finite integral and the corresponding
IR-divergent on-shell integral,
\beqar\label{eq:rc0diff}
\lefteqn{\Cbr_0(p_1,p_2,0,m_1,m_2) - 
\Cbr_0\Bigl(p_1,p_2,\la,\sqrt{p_1^2},\sqrt{p_2^2}\Bigr)} \nn\\
&=& \int\!\frac{\rd^3{\bf q}}{\pi q_0}\,
\biggl\{\frac{1}{(2p_1q+p_1^2-m_1^2)(2p_2q+p_2^2-m_2^2)}
-\frac{1}{(2p_1q)(2p_2q)}\biggr\}\bigg|_{q_0=\sqrt{{\bf q}^2+\lambda^2}},
\eeqar
which is UV-finite and Lorentz-invariant. Extracting the signs 
$\sigma_i$ of $p_{i0}$ via definition \refeq{eq:momredef}, and assuming
for the moment $\sigma_i\Re(p_i^2-m_i^2)>0$, the Feynman-parameter
representation \refeq{eq:FeynPar} can be applied to each term in the
integrand of \refeq{eq:rc0diff}. The integration over $\rd^3{\bf q}$ 
can be carried out and yields 
\beqar\label{eq:rc0diffFP}
\lefteqn{\Cbr_0(p_1,p_2,0,m_1,m_2) -
\Cbr_0\Bigl(p_1,p_2,\la,\sqrt{p_1^2},\sqrt{p_2^2}\Bigr)} \\
&=& \sigma_1\sigma_2\int_0^\infty\rd x_1\rd x_2\,
\de\biggl(1-\sum_{i=1}^2\alpha_i x_i\biggr)
\frac{\disp \ln(\lambda^2)
+\ln\biggl[\biggl(\sum_{i=1}^2\tilde p_i x_i\biggr)^2\biggr]
-2\ln\biggl[\sum_{i=1}^2\sigma_i(p_i^2-m_i^2)x_i\biggr]}
{\disp 2\biggl(\sum_{i=1}^2\tilde p_i x_i\biggr)^2}. \nn
\eeqar
Putting for instance $\al_1=0$ and $\al_2=1$, and using
\beq
\int_0^\infty \rd x \left(\frac{1}{x+z_1}-\frac{1}{x+z_2}\right)
\ln(1+z x)=
\cLi(z_1,z)-\cLi(z_2,z), \nl
\eeq
where $|\arc(z_{1,2},z,z\,z_{1,2})|<\pi$,
the remaining one-dimensional integration and the analytical
continuations to arbitrary complex $(p_i^2-m_i^2)$ are straightforward.

Next we consider the IR-finite 4-point function
\beqar
\lefteqn{\Dbr_0(p_1,p_2,p_3,0,m_1,m_2,m_3)} \nn\\
&=& \left.\int\!\frac{\rd^{3}{\bf q}}{\pi q_0}\,
\frac{1}{(2p_1q+p_1^2-m_1^2)(2p_2q+p_2^2-m_2^2)(2p_3q+p_3^2-m_3^2)}
\right|_{q_0=|{\bf q}|}.
\eeqar
For $\sigma_i\Re(p_i^2-m_i^2)>0$ we can proceed as above and find
\beq\label{eq:rd0diffFP}
\Dbr_0(p_1,p_2,p_3,0,m_1,m_2,m_3) 
= \sigma_1\sigma_2\sigma_3\int_0^\infty\rd x_1\rd x_2\rd x_3\,
\frac{\disp \de\biggl(1-\sum_{i=1}^3\alpha_i x_i\biggr)}
{\disp \biggl[\sum_{i=1}^3\sigma_i(p_i^2-m_i^2)x_i\biggr]
\biggl(\sum_{i=1}^3\tilde p_i x_i\biggr)^2}.
\eeq
Again, the two-dimensional integration over the Feynman parameters
and the analytical continuations in $(p_i^2-m_i^2)$ are straightforward.

Finally, we inspect the IR-divergent 4-point function
\beq
\Dbr_0\Bigl(p_1,p_2,p_3,\la,\sqrt{p_1^2},m_2,\sqrt{p_3^2}\Bigr) =
\left.\int\!\frac{\rd^{3}{\bf q}}{\pi q_0}\,
\frac{1}{(2p_1q)(2p_2q+p_2^2-m_2^2)(2p_3q)}
\right|_{q_0=\sqrt{{\bf q}^2+\la^2}}.
\eeq
Instead of applying the Feynman-parameter representation directly, it is
more convenient to extract the IR singularity by subtracting and adding
an IR-divergent 3-point function containing the same IR structure. Since
the difference between
the 4- and 3-point functions is IR-finite, we can
regularize the IR divergence in these two integrals in a more convenient
way. Therefore, we write
\beqar
\lefteqn{\Dbr_0\Bigl(p_1,p_2,p_3,\la,\sqrt{p_1^2},m_2,\sqrt{p_3^2}\Bigr)
= \lim\limits_{m^2\rightarrow p_1^2}
\bigg[\Dbr_0\Bigl(p_1,p_2,p_3,0,m,m_2,\sqrt{p_3^2}\Bigr)}
\nn\\
&& 
-\frac{1}{p_2^2-m_2^2}\Cbr_0\Bigl(p_1,p_3,0,m,\sqrt{p_3^2}\Bigr)
+\frac{1}{p_2^2-m_2^2}\Cbr_0\Bigl(p_1,p_3,\la,\sqrt{p_1^2},\sqrt{p_3^2}\Bigr)
\bigg],
\eeqar
\ie we regularize the IR divergence in the 4-point function by the
off-shellness $p_1^2-m^2\ne 0$. Both the 4-point function as well as the
difference of 3-point functions can be treated as above, yielding
straightforward Feynman-parameter integrals according to 
\refeq{eq:rc0diffFP} and \refeq{eq:rd0diffFP}, respectively. 
The limit $m^2\to p_1^2$
can easily be taken after the integrations have been performed.

\section{Explicit results for scalar integrals}
\label{appscalint}

\subsection{Loop integrals in double-pole approximation}
\label{appvirtual}

In the (\mmp) correction the following
combination of 3-point functions appears in the strict DPA:
\beqar\label{dC0virt}
\lefteqn{ C_0(k_+,-k_-,0,M,M) -
\Bigl[C_0(k_+,-k_-,\la,\MW,\MW)\Bigr]_{k_\pm^2=\MW^2} }
\quad\nn\\
&\sim& \frac{1}{s\beta_\PW}\biggl\{
-\cLi\biggl(\frac{K_-}{K_+},x_\PW\biggr)
+\cLi\biggl(\frac{K_-}{K_+},x_\PW^{-1}\biggr)
+\Li(1-x_\PW^2)+\pi^2 +\ln^2(-x_\PW)
\nn\\ && {}
+2\ln\biggl(\frac{K_+}{\lambda\MW}\biggr)\ln(x_\PW) -2\pi\ri\ln(1-x_\PW^2)
\biggr\}.
\eeqar
In order to include the full off-shell Coulomb singularity, one has to
add
\beq
 \frac{2\pi\ri}{s \betap} \ln\biggl(
    \frac{\betaM+\Delta_M-\betap}{\betaM+\Delta_M+\betap}\biggr) 
-  \frac{2\pi\ri}{s \betaW} \ln\biggl(
    \frac{K_+ + K_- + s\betaW\Delta_M}{2\betaW^2s}\biggr) 
\eeq
to the r.h.s.\ of \refeq{dC0virt}.

For the (\mfp) and (\ffp) corrections we need the following 4-point functions:
\beqar \label{D00}
\lefteqn{ D_0(0) = D_0(-k_4,k_+ +k_3,k_2+k_3,0,M,M,0) }
\quad\nn\\
&\sim& \frac{1}{\kappa_\PW} \sum_{\si=1,2} (-1)^\si \biggl\{
 \cLi\biggl(-\frac{s_{13}+s_{23}}{\MW^2}-\ri\epsilon,-x_\si\biggr)
+\cLi\biggl(-\frac{\MW^2}{s_{23}+s_{24}}+\ri\epsilon,-x_\si\biggr)
\nn\\ && \qquad {}
-\cLi\biggl(x_\PW,-x_\si\biggr)
-\cLi\biggl(x_\PW^{-1},-x_\si\biggr)
-\ln\biggl(1+\frac{s_{24}}{s_{23}}\biggr)\ln(-x_\si)
\biggr\},
\\[.5em] &&
\mbox{with} \qquad
x_1 = \frac{s_{24}z}{\MW^2}, \quad x_2 = \frac{\MW^2}{s_{13}z}, \quad
z = \frac{\MW^4+s_{13}s_{24}-s_{14}s_{23}+\kappa_\PW}{2s_{13}s_{24}},
\label{eq:xtilde}
\\[1em]
\lefteqn{ D_0(1) = D_0(-k_-,k_+,k_2,0,M,M,m_2) }
\quad\nn\\
&\sim& \frac{1}{K_+(s_{23}+s_{24})+K_-\MW^2} \biggl\{
\sum_{\tau=\pm 1} \biggl[
 \cLi\biggl(\frac{K_+}{K_-},x_\PW^\tau\biggr)
-\cLi\biggl(-\frac{\MW^2}{s_{23}+s_{24}}+\ri\epsilon,x_\PW^\tau\biggr)
\biggr]
\\ && {}
-2\cLi\biggl(\frac{K_+}{K_-},-\frac{s_{23}+s_{24}}{\MW^2}-\ri\epsilon\biggr)
-\ln\biggl(\frac{m_2^2}{\MW^2}\biggr) \biggl[
\ln\biggl(\frac{K_+}{K_-}\biggr)
+\ln\biggl(-\frac{s_{23}+s_{24}}{\MW^2}-\ri\epsilon\biggr) \biggr]
\biggr\},
\hspace{1.8em}
\nn\\[1em]
\lefteqn{ D_0(2) = D_0(-k_3,k_+,k_2,\la,m_3,M,m_2) }\quad
\nn\\*
&\sim& \frac{1}{K_+s_{23}} \biggl\{
-\Li\biggl(-\frac{s_{13}}{s_{23}}\biggr)
-\frac{\pi^2}{3}
+2\ln\biggl(-\frac{s_{23}}{m_2 m_3}-\ri\epsilon\biggr)
\ln\biggl(\frac{-K_+}{\la\MW}\biggr)
-\ln^2\biggl(\frac{m_2}{\MW}\biggr)
\nn\\ && {}
-\ln^2\biggl(-\frac{s_{13}+s_{23}}{m_3\MW}-\ri\epsilon\biggr)
\biggr\},
\\[1em]
\lefteqn{D_0(3) = D_0(-k_3,-k_-,k_2,\la,m_3,M,m_2) 
= D_0(2) \Big|_{K_+\leftrightarrow K_-,m_2\leftrightarrow m_3,
s_{13}\leftrightarrow s_{24}},
}\quad
\\[1em]
\lefteqn{ D_0(4) = D_0(-k_3,-k_-,k_+,0,m_3,M,M) 
= D_0(1) \Bigr|_{K_+\leftrightarrow K_-,m_2\leftrightarrow m_3,
s_{13}\leftrightarrow s_{24}}.
}\quad
\eeqar

\subsection{Bremsstrahlung integrals in double-pole approximation}
\label{appreal}

In the (\mmp) interference corrections the following
combination of 3-point functions appears:
\beqar\label{C0br}
\lefteqn{ \Cbr_0(k_+,k_-,0,M,M^*) -
\Bigl[\Cbr_0(k_+,k_-,\la,\MW,\MW)\Bigr]_{k_\pm^2=\MW^2} }
\quad\nn\\
&\sim& 
\biggl\{C_0(k_+,-k_-,0,M,M)- \Bigl[
C_0(k_+,-k_-,\la,\MW,\MW)\Bigr]_{k_\pm^2=\MW^2}\biggr\}\bigg|_{K_-\to -K_-^*}
\nn\\ && {}
- \frac{2\pi\ri}{s\beta_\PW}
\ln\biggl[\frac{K_+ +K_-^*x_\PW}{\ri\lambda\MW(1-x_\PW^2)}\biggr].
\eeqar

For the (\mfp) and (\ffp) interference corrections the
following 4-point functions are required:
\beqar\label{Dbr00}
\lefteqn{\tilde\Dbr_0(0) = -D_0(k_4,k_+-k_3,k_2-k_3,0,M_-,M_+,0)  } \quad&&
\nn\\
&\sim& -D_0(0)
+\frac{2\pi\ri}{\kappa_\PW}\biggl\{
\ln\biggl(-x_\PW\frac{s_{23}}{\MW^2}\biggr)
+\ln\biggl[1+\frac{s_{13}}{s_{23}}(1-z)\biggr]
+\ln\biggl[1+\frac{s_{24}}{s_{23}}(1-z)\biggr] 
\nn\\ && {} 
\qquad\qquad
-\ln\biggl(1+\frac{s_{13}}{\MW^2}z x_\PW\biggr)
-\ln\biggl(1+\frac{s_{24}}{\MW^2}z x_\PW\biggr) \biggr\},
\\[.5em] && 
\mbox{with $z$ from \refeq{eq:xtilde},} 
\nn\\[1em]\label{D0br1}
\lefteqn{ \Dbr_0(1) = \Dbr_0(k_-,k_+,k_2,0,M^*,M,m_2) }\quad
\nn\\
&\sim& D_0(1)\Big|_{K_-\to -K_-^*}
+\frac{2\pi\ri}{K_+(s_{23}+s_{24})-K_-^*\MW^2} \biggl[
2\ln\biggl(1-\frac{K_+}{K_-^*}\frac{s_{23}+s_{24}}{\MW^2}\biggr)
\nn\\ && {}
\qquad\qquad\qquad\qquad
-\ln\biggl(1+\frac{K_+}{K_-^* x_\PW}\biggr)
-\ln\biggl(1+\frac{x_\PW \MW^2}{s_{23}+s_{24}}\biggr)
-\ln\biggl(\frac{m_2^2}{\MW^2}\biggr) 
\biggr],
\hspace{2em}
\\[1em]
\lefteqn{ \Dbr_0(2) = \Dbr_0(k_3,k_+,k_2,\la,m_3,M,m_2) }\quad
\nn\\
&\sim& D_0(2)
+\frac{2\pi\ri}{K_+s_{23}} 
\ln\biggl[\frac{K_+s_{23}}{\ri\lambda m_2(s_{13}+s_{23})}\biggr],
\\[1em]
\lefteqn{\Dbr_0(3) = \Dbr_0(k_3,k_-,k_2,\la,m_3,M^*,m_2) 
}\quad
\nn\\
&\sim& -D_0(3)\Big|_{K_-\to -K_-^*}
+\frac{2\pi\ri}{K_-^*s_{23}}
\ln\biggl[\frac{\ri K_-^*m_2}{\lambda(s_{23}+s_{24})}\biggr],
\\[1em]
\lefteqn{ \Dbr_0(4) = \Dbr_0(k_3,k_-,k_+,0,m_3,M^*,M) 
}\quad
\nn\\
&\sim& -D_0(4)\Big|_{K_-\to -K_-^*}
+\frac{2\pi\ri}{K_-^*(s_{13}+s_{23})-K_+\MW^2}
\biggl[\ln\biggl(1+\frac{K_-^* x_\PW}{K_+}\biggr)
-\ln\biggl(1+\frac{x_\PW \MW^2}{s_{13}+s_{23}}\biggr)\biggr].
\nn\\
\eeqar
We note that the logarithmic terms on the r.h.s.\ of \refeq{Dbr00}
yield purely imaginary contributions if
$(s_{13}+s_{23})>-\MW^2 x_\PW$, $(s_{23}+s_{24})>-\MW^2 x_\PW$, 
and $\kappa_\PW$ is imaginary. These conditions are fulfilled on 
resonance. Off resonance, this is no longer true
near the boundary of phase space.
But since these conditions are only violated in a fraction of
phase space of order $|k_\pm^2-\MW^2|/\MW^2$, which is irrelevant in DPA
(compare the discussion of the relevance of the Landau singularities in
\refse{se:ambig}), it would
even be allowed to replace $\tilde\Dbr_0(0)$, which is real,
by $-\Re\{D_0(0)\}$, as it was done in \citeres{Be97a,Be97b}.

\end{document}